\include{graphicsx} 

\documentclass[12pt,preprint]{aastex}

\slugcomment{Accepted by ApJ on June 21, Published Aug 20, 2013}

\shorttitle{Visible AO observations of the Trapezium Cluster}
\shortauthors{Close et al.}

\begin{document}


\title{Diffraction-limited Visible Light Images of Orion Trapezium Cluster With the Magellan Adaptive Secondary AO System (MagAO){\footnotemark}}

\footnotetext{$^1$This paper includes data gathered with the 6.5 meter Magellan Telescopes located at Las Campanas Observatory, Chile. }


\author{Close, L.M.$^1$, Males, J.R.$^1$, Morzinski, K.$^{1,4}$, Kopon, D.$^{1,5}$, Follette, K.$^1$, Rodigas, T.J.$^1$, Hinz, P.$^1$, Wu, Y-L. $^1$, Puglisi, A.$^{2}$, Esposito, S.$^2$, Riccardi, A.$^2$, Pinna, E.$^2$, Xompero, M.$^2$, Briguglio, R.$^2$, Uomoto, A, $^3$, Hare, T. $^3$
}

\email{lclose@as.arizona.edu}

\affil{$^1$Steward Observatory, University of Arizona, Tucson, AZ 85721, USA}
\affil{$^2$INAF - Osservatorio Astrofisico di Arcetri, I-50125, Firenze, Italy}
\affil{$^3$Carnegie Observatories, 813 Santa Barbara Street, Pasadena, California, 91101 USA}
\affil{$^4$NASA Sagan Fellow}
\affil{$^5$MPiA, Germany}



\begin{abstract}

We utilized the new high-order (250-378 mode) Magellan Adaptive Optics system (MagAO)
to obtain very high spatial resolution observations in ``visible light'' with MagAO's
VisAO CCD camera. In the good-median seeing conditions of Magellan ($0.5-0.7\arcsec$)
we find MagAO delivers individual short exposure images as good as 19 mas
optical resolution. Due to telescope vibrations, long exposure (60s) r' (0.63$\mu m$) images are slightly
coarser at FWHM=23-29 mas (Strehl $\sim28\%$) with bright ($R<9$ mag) guide stars. These are the highest resolution filled-aperture images
published to date.  Images of the young ($\sim 1$ Myr) Orion Trapezium
$\theta^{1}$ Ori A, B, and C cluster members were obtained with
VisAO. In particular, the 32 mas binary $\theta^{1}$ Ori $C_1C_2$ was
easily resolved in non-interferometric images for the first time. Relative
positions of the bright trapezium binary stars were measured with
$\sim 0.6-5$ mas accuracy. We now are
sensitive to relative proper motions of just $\sim0.2$ mas/yr
($\sim0.4$ km/s at 414 pc) -- this is a $\sim2-10\times$ improvement in
orbital velocity accuracy compared to previous efforts.
For the first time, we see clear motion of the barycenter of
$\theta^{1}$ Ori $B_{2}B_{3}$ about $\theta^{1}$ Ori
$B_{1}$. All five members of the $\theta^{1}$ Ori $B$ system appear
likely a gravitationally bound ``mini-cluster'', but we find that not
all the orbits can be both circular and co-planar. The lowest mass
member of the $\theta^{1}$ Ori $B$ system ($B_{4}$; mass $\sim 0.2
M_{\sun}$) has a very clearly detected motion (at $4.1\pm1.3$ km/s;
correlation=99.9\%) w.r.t $B_1$. Previous work has suggested that $B_4$
and $B_3$ are on long-term unstable orbits and will be ejected from
this ``mini-cluster''.  However, our new ``baseline'' model of the
$\theta^{1}$ Ori $B$ system suggests a more hierarchical system than
previously thought, and so the ejection of $B_4$ may not occur for
many orbits, and $B_3$ may be stable against ejection long-term. This
``ejection'' process of the lowest mass member of a ``mini-cluster'' could play a
major role in the formation of low mass stars and brown dwarfs.

\end{abstract}


\section {INTRODUCTION}

\subsection{The Need For High-Resolution Imaging}

It is critical to the understanding of the motions and masses of stars, brown
dwarfs, and exoplanets to obtain the highest resolution images
possible. In fact, almost every aspect of astronomical science benefits from the highest spatial resolutions possible. The highest resolution ``maps'' at the milliarcsec (mas) resolution scales ($0.001\arcsec$ = 1 mas) are 
being produced by interferometry (like VLTI/AMBER in the IR). However,
interferometric techniques suffer from incomplete uv coverage and
object models are usually required to interpret interferometric
data. Moreover, combining multiple 8m telescopes together in the VLTI
and waiting for the Earth's rotation to fill in the uv plane is
both time consuming and expensive (hence limiting the general
utility of large surveys with VLTI, for example). 

Imaging from space with a filled aperture (and so complete uv coverage) with {\it HST} has proven to be very
productive, but {\it HST}'s small 2.4m aperture, combined with a need for
large pixels, limits its best spatial resolutions to 50-100
mas. Also {\it HST} is considerably more expensive than any other
telescope and its lifetime is limited. Large (8-10m) ground-based
telescopes can match HST's $\sim50$ mas V-band resolution with
adaptive optics in the NIR (1.2-2.4 $\mu m$). For example, the 8.4m
LBTAO system (FLAO; \cite{esp11}) can also achieve deep 50 mas
resolution images with AO at 1.64 $\mu m$ \citep{clo12b}. However, to
achieve deep images better than $\sim40-50$ mas is
impossible with the current generation of facility AO systems and
cameras. For example, to reach $\le20$ mas resolutions at H band (1.65 $\mu m$) would require a
D$\ge16$m filled-aperture telescope. Hence it will not be until the ELT era (early, to mid, 2020s) that images in the NIR
will be significantly sharper than 20 mas.

\subsection{Into the Blue: Adaptive Optics in the Visible}

However, there is another approach to reaching these
resolutions. While 8-10m AO performance is limited to $\ge 40$ mas in
the NIR, it is possible to gain a factor of two improvement in resolution
by moving to shorter (bluer) wavelengths for AO correction. This so
called ``visible AO'' can theoretically reach 16 mas resolutions on an
8m telescope at 0.656 $\mu m$ ($H_\alpha$). However, the complexity of
an 8-10m class AO system designed for optical wavelengths ($>500$
modes, at $>1$ KHz) is beyond that of the current facility systems
(with perhaps the exception of the FLAO system on the 8.4m LBT which; however, currently has no facility visible AO CCD science
camera \citep{esp10a}). 

We note that AO with ``lucky'' imaging in the visible has been
successfully used at the somewhat smaller 5m Palomar \citep{law09} and
has reached resolutions of 35 mas, recently the Palm3000 system has demonstrated excellent corrections \citep{dek13}. Improved Lucky visible imaging (image synthesis based on Fourier Amplitude selection) has also been developed by \cite{gar12}. Visible AO has been done before on much smaller telescopes like
Robo-AO on the 1.5 m at Palomar \citep{bar12} or the Villages project
on the 1.0m Nickel at Lick \citep{mor10}. In the near future some
polarization work will be done in the visible with the 8m VLT with the
SPHERE AO system and ZIMPOL \citep{baz12}.  Yet, MagAO is the first large ($D\ge6.5$ m)
telescope AO system designed to work in the visible --complete with a
facility CCD AO science camera (VisAO). The MagAO commissioning
results presented here inform us on the utility of large telescope visible AO performance.

\subsection{The Magellan AO System}

We have developed an AO system (inspired, in large part, by
LBT's FLAO system; \cite{esp12}) that can reach 20 mas resolutions with just 250-378
modes at 1 kHz sampling speeds. It is important that such a visible AO
system be located at an excellent site where the median seeing is less
than $0.64\arcsec$. To achieve an AO fitting error small enough to
reach $110$ nm rms total wavefront error (WFE) with 250 modes
requires a telescope diameter of D$\le6.5$m. Hence, a solution to
this design problem is a fast ($<1$ ms response time) 585 element
second generation adaptive secondary mirror (ASM) with a 1 kHz Pyramid
wavefront sensor (PWFS). These are exactly the characteristics of the
Magellan Adaptive Secondary AO system (MagAO) deployed on the
$0.64\arcsec$ median seeing \citep{tho10} 6.5m Magellan telescope at Las Campanas
Observatory, Chile.

MagAO, with its VisAO camera\footnotemark, is the first large telescope ($\ge6.5$m) facility AO system
\footnotetext{see {\it http://visao.as.arizona.edu/} for web resources, SPIE publications, and guides for observers}
deployed that is targeting science in the visible (0.6-1.1 $\mu m$). As will be shown in later in this paper, MagAO at first light produced long exposure (60s) diffraction-limited (110 nm WFE; $28\%$ Strehl) 0.63$\mu m$ images. Note during the first light run (commissioning run \#1; Nov/Dec 2013) we were limited to 250 corrected modes\footnotemark. For more technical details about MagAO itself
please see \cite{clo12a}.

It is important to note that MagAO sends all the infrared light into the
Clio2 NIR (1-5.3 $\mu m$) camera \citep{hin10, mor13}), whereas the visible
light ($\lambda<1.1\mu m$) is split by a selectable beamsplitter
between the PWFS and the VisAO (0.6-1.1$\mu m$) science camera (for more on the VisAO camera see \cite{mal12, kop12, clo12a}). Hence all three focal stations (Clio2, VisAO, and PWFS) work simultaneously on all targets, allowing Visible and IR science to be done simultaneously.

\footnotetext{During the recent commissioning run \#2 (March/April 2013) MagAO had a better modal basis set which allowed on-sky closed-loop stability at its maximum of 378 corrected modes. Hence MagAO currently can achieve a WFE of just 102 nm rms with 378 modes in $0.5\arcsec$ seeing on bright stars ($R\le9$ mag). However, this paper concerns the first light (commissioning run $\#1$) results of MagAO where only 250 modes were corrected.}

\subsection{First Light VisAO Science: Motions of the Massive Young Stars in The Orion Trapezium Cluster}

Clearly the exciting possibility of obtaining $\sim20$ mas FWHM images with MagAO could enhance our understanding of the positions (and motions) of the nearest massive young stars. Hence we targeted the Orion Trapezium cluster during the first light commissioning run with the MagAO system. 

The study of the motions of the young stars in the Trapezium cluster is an important problem (see for example \cite{mcc03,clo12b,gil13}). After all,    
the detailed formation of stars is still a poorly understood process. In
particular, the formation mechanism of the lowest mass stars and brown dwarfs is
uncertain. Detailed 3D (and N-body) simulations of star formation by \cite{bat02,bat03,bat09,bat11} and \cite{par11} all
suggest that stellar embryos frequently form into ``mini-clusters'' which
dynamically decay, ``ejecting'' the lowest mass members. Such theories
can explain why there are far more field brown dwarfs (BD) compared to
BD companions of solar type stars \citep{mcc03} or early M stars
\citep{hin02}. Moreover, these theories which invoke some sort of
dynamical decay \citep{dur01} or ejection \citep{rep02} suggest that
there should be no wide ($>20$ AU) very low mass (VLM; $M_{tot}<0.185
M_{\sun}$) binary systems observed in the field (age $\sim 5$ Gyr). Indeed, the AO surveys of
\cite{clo03a} and the HST surveys of \cite{rei01a,bur03,bou03,giz03} have not
discovered more then a few wide ($>16$ AU) VLM systems in the field population (for a review see \cite{bur07}). Additionally, the dynamical biasing towards the ejection of the
lowest mass members naturally suggests that the frequency of field VLM
binaries should be much lower ($\la 5\%$ for $M_{tot}\sim 0.16
M_{\sun}$) than for more massive binaries ($\sim 60\%$ for $M_{tot}
\sim 1 M_{\sun}$). Indeed, observations suggest that the binarity of
VLM systems with $M_{tot} \la 0.185 M_{\sun}$ is $10-15\%$
\citep{clo03a, bur03, bur07} which, although higher than predicted is still
lower than that of the $\sim42\%$ of more massive M-dwarfs \cite{fis92} or $\sim 60\%$ of G star binaries \cite{duq91}. However as is noted in \cite{clo07} there is evidence that in young clusters wide VLM binaries are much more common than in the old field population. They attribute this to observing these wide VLM systems before they are destroyed by encounters in their natal clusters. Hence, we need to look at nearby young clusters to see these low-mass objects in ``mini-clusters'' (of a few bound stars) before ejection has occurred.  

Despite the success of these decay, or ejection, scenarios in predicting
the observed properties of low mass VLM stars and binaries, it is still not clear that
``mini-clusters'' even exist in the early stages of star formation. To
better understand whether such ``mini-clusters'' do exist we have
examined the closest major OB star formation cluster for signs of such
``mini-clusters''. Here we focus on the $\theta^1$ Ori stars in the
famous Orion Trapezium cluster. Trying to determine if some of the tight star
groups in the Trapezium cluster are gravitationally bound is a first
step to determining if bound ``mini-clusters'' exist. Also it is
important to understand the true number of real, physical, binaries in
this cluster, as there is evidence that the overall number of binaries
is lower (at least for the lower mass members) in the dense trapezium
cluster compared to the lower density young associations like
Taurus-Auriga \citep{mcc11,koh06}.  In particular, we will examine the
case of the $\theta ^1$ Ori A, B and C groups in detail.

\begin{figure}
\includegraphics[angle=0,width=\columnwidth]{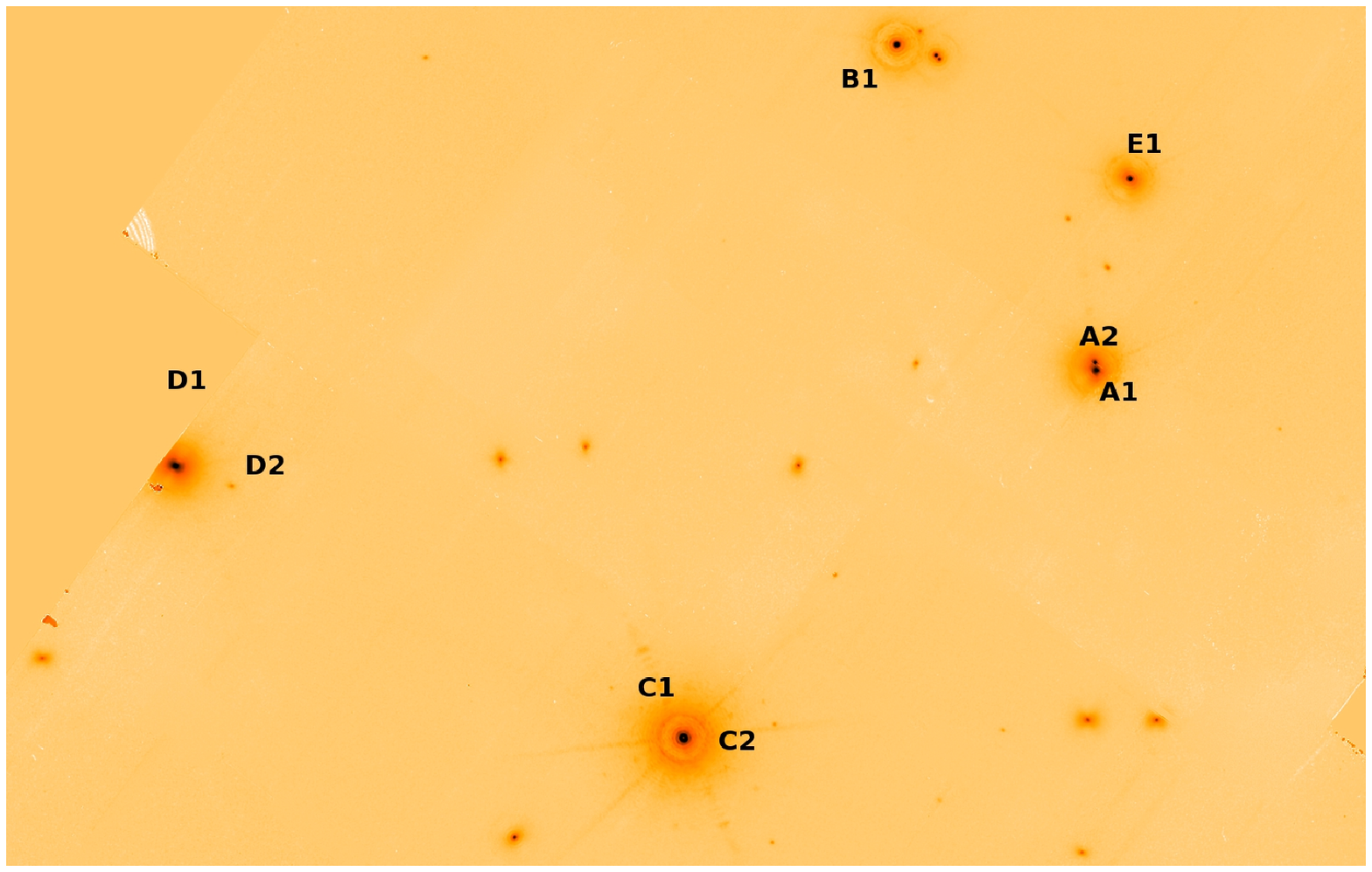}\caption{
The locations and nomenclature of \cite{clo03c} of the $\theta^{1}$ Ori Trapezium stars as imaged over $\sim35\times 30\arcsec$ FOV at the LBT with LBTAO/PISCES in [FeII] (reproduced from \cite{clo12b}). Logarithmic color scale. North is up and east is left. Note that the object ``$A_1$'' is really a spectroscopic binary ($A_1A_3$); where the unseen companion $A_3$ is separated from $A_1$ by 1 AU \citep{bos89}. The B group is shown in more detail in Figs. \ref{fig4} - \ref{mag_B}. It is not currently clear if $D_2$ is physically related to $D_1$. $E_1$ appears to be a single star. No new faint companions were discovered (at $>5\sigma$) around any of the Trapezium stars in this study.
}
\label{fig2}
\end{figure}

The Trapezium OB stars ($\theta^1$ Ori A, B, C, D, and E; see Fig. \ref{fig2}) consists of
the most massive OB stars located at the center of the Orion Nebula
star formation cluster (for a review see \cite{gen89}). Due to the
nearby (VLBA trigonometric parallax distance of $414\pm7$ pc; \cite{men07}),
and luminous nature of these stars they are a unique laboratory for the study of a high-mass star formation cluster (the dominate birthplace for stars of all masses), and have been the target of several
high-resolution imaging studies. For brevity, we do not reproduce here a complete history of past high resolution surveys of Trapezium, please see \cite{clo12b} instead for a review.

\cite{clo12b} utilized the LBT FLAO system to map
out the Trapezium in narrow-band NIR filters at $\sim50-60$ mas
resolutions. It total \cite{clo12b} analyzed 14 years of observations of the
cluster. Yet, only the LBT 2011 observations were of very high
quality. In this paper we present the first high-resolution visible
(0.57-0.68 $\mu m$) AO images. These images are of the Trapezium
cluster and reach very high resolutions of $\sim23$ mas. We have now over 15
years of observations of this field with at $<0.08\arcsec$
resolution. More importantly, we now have two complete high-quality
datasets from LBT and MagAO, that track motion of the
Trapezium stars at $<0.05\arcsec$ resolutions.

In this paper we outline how these MagAO observations were carried out
with the new VisAO camera. We detail
how these data were calibrated and reduced and how the stellar
positions were measured. We resolve the 32 mas binary $\theta^1$ Ori C in a filled aperture image for the first time. We compare the measured astrometry for $\theta^1$ Ori C1 and C2 against its published (interferometric) orbit. We also fit the observed positions to calculate
velocities (or upper limits) for the $\theta^1$ Ori $B_1,B_2,B_3,B_4$ \& $A_1,A_2$ stars.
While \cite{sch03} and \cite{clo03c, clo12b} had hints that the $\theta^1$ Ori
B group may be a bound ``mini-cluster'' ---we here show it is clearly so,
with the first detection of curvature in the orbital motion of members of this group. We also present the first model for how the complex set of orbits in the $\theta^1$ Ori B mini-cluster could (and cannot) be arranged.  

\section{INSTRUMENTAL SET-UP}

 We utilized MagAO to obtain the first diffraction-limited (and  unsaturated) images of the young
 stars in the Trapezium cluster in the visible ($0.6-0.7\mu m$). This is not a
simple task, since as telescopes have increased in size, bright stars tend to now saturate --even in the shortest possible exposures. Hence special precautions are needed to avoid saturation of the bright Trapezium stars themselves. It is difficult to make unsaturated, but diffraction-limited, ``visible light'' images of the bright Trapezium stars with modern 6.5m class AO systems at even moderately high Strehl. Witness the fact that this is the first such dataset ever published. The following subsections outline how this was accomplished.

The MagAO system is unique (at least in the southern hemisphere) in many ways. To reduce
the aberrations caused by atmospheric turbulence all large telescope AO systems have a
deformable mirror which is updated in shape at $\sim 500$ Hz. Except for the MMTAO and LBTAO systems \citep{wil03,esp11} 
all other adaptive optics systems have located this deformable mirror
(DM) at a re-imaged pupil (effectively a compressed image of the
primary mirror). To reimage the pupil onto a DM typically requires {\bf 3-8}
warm additional optical surfaces, which significantly increases the
thermal background and decreases the optical throughput of the system
\citep{llo00}. However, MagAO utilizes a next generation adaptive secondary DM.
 This DM is both the secondary mirror of the
telescope and the DM of the AO system. 
 In this
manner there are no additional optics required in front of the science
camera. Hence the emissivity is lower, throughput higher. MagAO's DM is an advanced ``second generation'' adaptive secondary mirror (ASM; similar to those on the LBT), which enables the highest on-sky visible Strehl ($>25\%$ at r' band; 0.57-0.68$\mu m$) of any large 6.5-10m telescope today. 

The MagAO ASM consists of 585 voice coil actuators that push (or pull) on 585 small
magnets glued to the backsurface of a thin (1.6 mm), 850 mm
aspheric ellipsoidal Zerodur glass ``shell'' (for a detailed review of the secondary mirror see \cite{clo12a}).
 Like in the case of the LBT AO system we have complete positional
control of the surface of this reflective shell by use of a 70kHz
capacitive sensor feedback loop. This positional feedback loop allows
one to position an actuator of the deformable shell to within $\sim5$
nm rms (total residual polishing wavefront errors (mainly at inter actuator scales) amount to only $\sim50$ nm rms over the whole secondary). The AO system samples (and drives the ASM)
at 990 Hz using 250-378 active controlled modes (with 585 actuators)
on bright stars (R$<9$ mag).\footnotemark

\footnotetext{With the PWFS we can operate on fainter stars by increasing the binning of the CCD39 in the PWFS. Hence for fainter guide stars with $9<R<12.7$ mag bin $2\times2$ and 120 modes are used. Likewise for $14.2<R<15.6$ mag bin $3\times3$ and 66 modes, $14.2<R<15.6$ mag bin $4\times4$ and 28 modes, and for the faintest stars ($15.6<R<16.5$ mag) then bin $5\times5$ and just 21 modes are corrected. Once we are fainter than $R>12.7$ mag visible AO correction is very poor and science can only be done in the NIR with Clio2.} 

The wavefront slopes are measured with a very accurate, well
calibrated, low aliasing error, Pyramid Wavefront sensor (PWFS). This
is the second large telescope to use a PWFS (after the
LBT, \cite{esp11}). The performance of the MagAO PWFS is
excellent. The very low residual wavefront errors obtained by the PWFS
+ ASM combination is due, in part, to the very accurate (high S/N)
interaction matrix that can be obtained in closed-loop daytime
calibrations with a retro-reflecting calibration return optic
(CRO; \cite{kop12}) that takes advantage of the Gregorian (concave)
nature of the secondary. To guarantee strict ``on-sky'' compliance
with the ``daytime calibrated'' interaction matrix pupil/ASM/PWFS
geometry the PWFS utilizes a novel ``closed-loop pupil alignment
system'' that maintains the pupil alignment to $<2.5\mu m$ (at the PWFS CCD39 images of the 4 pupils produced by the PWFS) during all
closed-loop operations on bright stars. Moreover, we use a fixed pupil mask
on the ASM to maintain the exact same pupil illumination when the CRO
is used and also when we are on sky -- so that our interaction matrices are
valid (on and off sky). For a detailed review of the MagAO system
see
\cite{clo12a} and references within. 

\subsection{The MagAO PSF and Calculating Strehl}

During the MagAO first light commissioning run we observed the
$\theta^{1}$ Ori A, B and C groups on the nights of Dec 3,4, and 8
2012 (UT). The AO system corrected the lowest 250 system modes and was
updated at 990 Hz. The PWFS pupil was close-loop stabilized and the
shell was protected from wind with a windscreen at the secondary
mirror. Cooling pumps (for Clio2, VisAO, and the PWFS) added some
vibrational blurring into the PSF. After commissioning run 1 these
pumps were much better isolated. Nevertheless, the PSFs were still
close to perfectly diffraction-limited. To better gauge the
effectiveness of the AO correction we need to be able to measure the long-exposure PSF and calculate the
Strehl of the PSF.

\begin{figure}
\includegraphics[angle=0,width=\columnwidth]{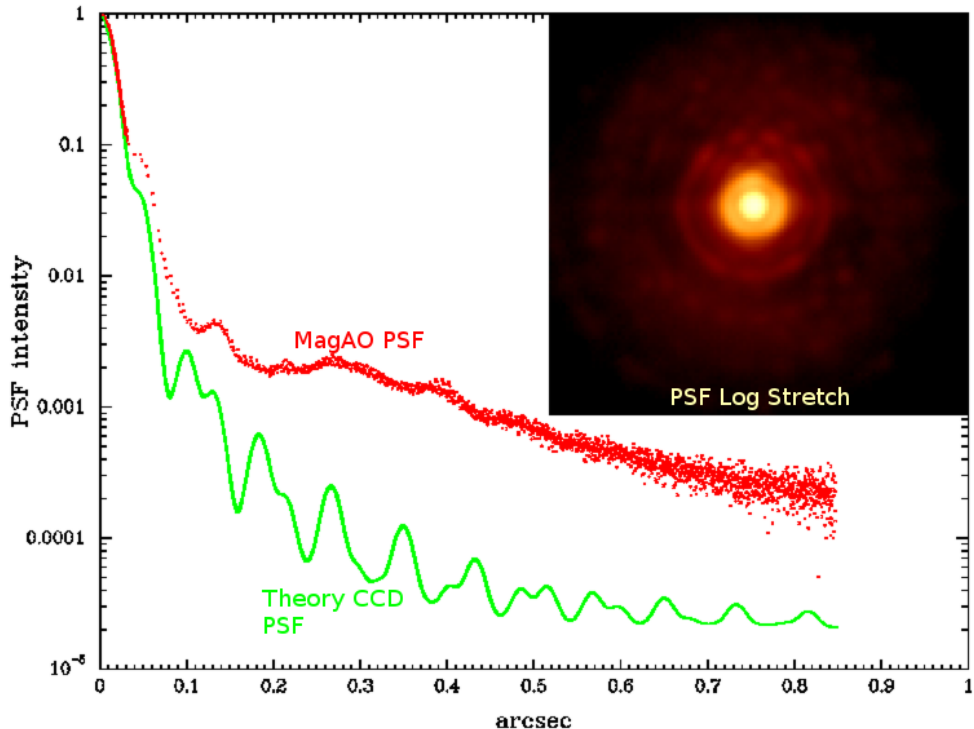}\caption{
 The radial profile {\it (red points)} of a deep (300s) MagAO $Y_{short}$ (0.98$\mu m$) PSF on a bright (V=5 mag) star closed-loop at 990 Hz with 250 modes in $0.6\arcsec$ V band seeing (from \cite{mal13}). {\it Insert:} a log10 Stretch of the PSF. There was no post-detection processing of any of the data (no SAA, no Lucky imaging or frame selection applied). The theoretical MagAO PSF profile as imaged by the E2V CCD47 (Strehl 100\%). A detailed comparison of the observed PSF to theory with our CCD47 (including dark current and PRF) shows that we reached a Strehl of 43\% or 140 nm rms optical wavefront error.       
}
\label{fig1}
\end{figure}

On bright ($R<9$) guide stars in $\sim 0.6\arcsec$ V-band seeing we could
obtain deep 5 minute PSF images (with no SAA or post-detection processing)
with Strehls of 43\% at $Y_{short}$ (Ys; 0.98$\mu m$), or 140 nm rm wavefront error (by use of the extended Marechal's approximation; see Fig. \ref{fig1}). We note the deep 5 minute image in Fig. \ref{fig1} suffered from some additional vibrational blurring due to the cooling pumps for the CCDs and Clio2\footnotemark. 
\footnotetext{For the data collection of Trapezium images in this paper the vibrating cooling pumps were temporarily powered off to help stabilize the images and obtain WFE $\sim$110 nm rms. We note that during the second commissioning run the Clio2 pump was successfully removed from the moving telescope structure and the CCD pump was better isolated from the telescope, greatly reducing residual vibrations.} 
These deep PSF images helped model the
PSF to calculate Strehls for MagAO on $\theta^{1}$ Ori C which was so
bright that only a $64\times 64$ CCD window could be readout without saturation on $C_1$. Hence the wings
of the PSF (beyond the $64\times 64$ window) had to be estimated from
a wavelength scaled PSF ``halo'' model based on the measured deep PSF wings of Figure \ref{fig1}. In this
manner realistic Strehls could be estimated reliably from the small
64x64 images of $\theta^{1}$ Ori $C_1$.\footnotemark We note that it was only the Strehl of $\theta^{1}$ Ori $C_1$ that required this bootstrap approach all other Strehls (from full frame CCD images) in this paper were measured in the usual manner by comparison to our model theoretical PSF.

\footnotetext{Note that to accurately calculate the Strehl of the $\theta^{1}$ Ori $C_1$ PSF required simply subtracting the PSF of $\theta^{1}$ Ori $C_2$ with the IRAF {\it daophot allstar} task.}   


\subsection{ The VisAO CCD AO Science Camera}

These observations utilized the first facility visible light AO science camera (VisAO; (\cite{mal12,kop12}). VisAO has a fast, frame transfer, 1024x1024 0.5-1.1 $\mu m$ E2V CCD47 detector. We
used the 64x64 window mode to minimize saturation on the array while observing $\theta^1$ Ori C1 (V=5.13 mag). While the $\sim10\times$ fainter $\theta^1$ Ori B1 (V=7.2) allowed the whole CCD to be readout without saturation.

The VisAO focal plane platescales were calibrated by the astrometry of
four stars in the HD 40887 quadruple system and $\theta^1$ Ori $B_1$ and $\theta^1$ Ori $E_1$
\footnote{Typically the stars in the Trapezium used for this platescale test move at only $\sim0.0015\arcsec$/yr so the platescale error over the $6.24\arcsec$ distance is $\sim2x10^{-4}$ error --- which is much smaller than the magnitude ($\sim 0.1\%$) of the platescale errors -- $\sim0.06\arcsec$ over this distance.} 
(see sections 3 and 4 for more details about how the images were first reduced).

The positions (found by the IRAF {\it allstar} PSF fitting task) of these stars from our VisAO images were compared to unsaturated
astrometry from \cite{clo12b} which itself is derived from the HST ACS astrometry of \cite{ric07}.  VisAO platescales and rms errors were then determined for the $H_\alpha$, [OI], r', i', z' and Ys filters with the IRAF {\it geomap} task. The platescale found was $0.0078513\pm0.000015\arcsec$/pix at $H_\alpha$ (0.656$\mu m$), and [OI] (0.63 $\mu m$) providing a $8.03\times 8.03\arcsec$ FOV with our f/52.5 beam on the CCD47's 13.0 $\mu m$ pixels. At
r' (0.63$\mu m$) the platescale was slightly coarser at
$0.007917\pm0.000015\arcsec$/pix, at z' (0.906$\mu m$) just slightly finer at 0.007911$\arcsec\pm$0.000012$\arcsec$/pix, and at $Y_{short}$ (0.982$\mu m$) 0.007906$\arcsec\pm0.000014\arcsec$/pix. By design, the f/16 beam (direct from the ASM) is slowed down to f/52.5 to yield these very fine 7.9 mas/pixel VisAO platescales. We note this is one of the finest platescales ever for a facility camera.

Small distortions were detected by dithering a binary across the VisAO
CCD. In this manner we detected a small change in the Y platescale
($\le1\%$) from the top of the array to the bottom. The exact formula to
correct a binary with a primary star at position X,Y of separation
$\delta x$ and $\delta y$ for any residual distortions is
$true_{\delta x} = measured_{\delta x} - \delta
dx/(abs[measured_{\delta x}]/110.0)$ and $true_{\delta y} =
measured_{\delta y} - \delta dy/(abs[measured_{\delta y}]/44.5)$ where
$\delta dx = -0.00038921676*(X-512) + 0.00084322443*(Y-512)$ and
$\delta dy = -0.00025760395*(X-512) - 0.0024045175*(Y-512)$ our
observations were near the center of the detector and so these
corrections were actually very small ($0.1-0.5\%$ or 0.1-5 mas changes to the $0.1-1\arcsec$ binaries), nevertheless
all binary observations in this paper have been fully distortion corrected.

To determine the orientation of the Y axis of the VisAO images (which were all taken with
the rotator following) it was first necessary to rotate each image counterclockwise (with the IRAF {\it rotate} task) by the ROTOFF FITS keyword value +90 degrees. At this point it was found by {\it geomap} that the direction of North was slightly ($0.890^{\circ}$)
East of VisAO's Y axis compared to the HST ACS \cite{ric07} and LBT images \citep{clo12b} of the field. Hence a final counterclockwise 
rotation of $-0.890^{\circ}$ was applied to the final image. The rms uncertainty adopted for the MagAO rotator angle is estimated as $\sim 0.3^{\circ}$ this is the maximum error seen between different images of these stars on different nights. We suspect that this value of $\sim 0.3^{\circ}$ is quite conservative based on the very low scatter in the PA fits shown later in this paper.

\section{OBSERVATIONS \& REDUCTIONS}

For the $\theta^{1}$ Ori C field we locked the AO system (at 990Hz, 250
modes) in $0.5-0.7\arcsec$ seeing on the bright O5pv binary star $\theta^{1}$ Ori $C$ (V=5.13 mag) and
used a $64\times 64$ window in the center of the VisAO CCD with a set of
$2608\times 0.023$ second (60s total) unsaturated exposures at $H_\alpha$, [OI], and r'. Immediately following the unsaturated
exposures a set of 60 second exposures were obtained with the AO off. We note that $\theta^{1}$ Ori C is really a $\sim0.03"$ binary composed of C1 and C2, (see \cite{kra07} for more details). 

Then the AO system was locked (250 modes, 990 Hz) on $\theta^{1}$ Ori B1
(V=7.96 mag) and VisAO was used over its fullframe ($1k\times 1k$) pixels to produce a
set of $212\times 0.283$s unsaturated (60s total) images at z'. 

 The individual frames
were reduced in a normal manner. We used our custom AO image
reduction script of \cite{clo03a} to sky/bias subtract,
cross-correlate (when needed), and median combine each image. The final individual image sets 
of the C and B fields each had a total exposure time of 1 min. Figure \ref{fig2} is a large FOV LBT NIR AO image from \cite{clo12b} that defines the nomenclature and relative positions of the Trapezium stars for clarity. 

In figure \ref{fig1} one can see the marked improvement in resolution ($\sim600$ mas to 34 mas) and Strehl ($\sim0.5\%$ to $43\%$) having the AO loop closed makes to a 300 second exposure. We note these Ys images are not post-detection ``frame selected'' (lucky imaging) nor shift and added (SAA) -- so they are true 300s open-shutter exposures.

\begin{figure}
\includegraphics[angle=0,width=\columnwidth]{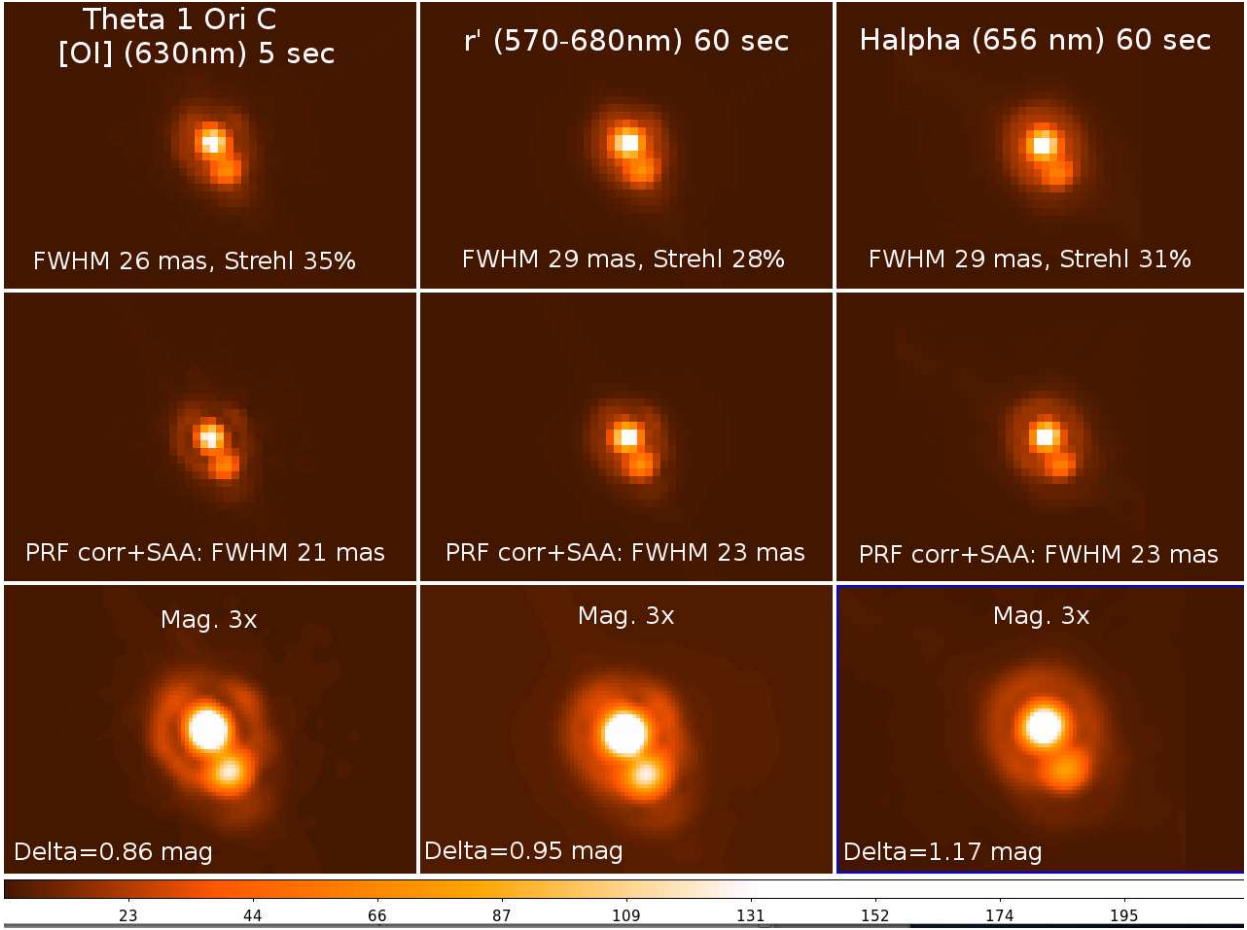}\caption{
{\it Top row:} the central ionizing binary of the Trapezium: $\theta^{1}$ Ori C as imaged with MagAO's VisAO CCD camera in different filters. Note the excellent resolution in the raw 60 second image. We note that no post-detection shift and add (SAA) was applied, nor was there any frame selection used to produce these top row images. Typically we achieved resolutions of $0.026-0.029\arcsec$ and Strehls of 28-35\% in $0.5-0.7\arcsec$ V-band seeing. {\it Middle row:} the same data as the top row, except the images have been post-detection aligned (SAA) and the pixel response function (PRF) has been removed. This improved image resolution by $\sim5-6$ mas. {\it Bottom row:} the row above is magnified by $3\times$ to better disply the data of the middle row. These are the highest resolution, deep, images ever obtained to our knowledge.
}
\label{fig3}
\end{figure}

In figure \ref{fig3} we show typical images of the binary $\theta^{1}$
Ori $C_1$ and $C_2$ imaged in $0.5\arcsec$ seeing ([OI] and r') on Dec 8, 2012 and worse
$0.7\arcsec$ seeing for $H_\alpha$ on Dec 3, 2012. In all cases
excellent (26-29 mas and 28-25\% Strehl) images are obtained. 

In the
middle row of figure \ref{fig3} we retrieve the true resolution of the
optical beam on the CCD47 by post-detection alignment of the images
($\sim2-4$ mas improvement). We also make a similarly small
resolution improvement ($\sim2-4$ mas) by removing the blurring
effects of the CCD47's pixel response function (PRF). The CCD47's
13.0$\mu m$ pixel PRF was calibrated by noting the slight improvement
in FWHM when a lab CCD with smaller $5.5\mu m$ pixels were used (instead of the $13 \mu m$ pixels) in a PRF lab test. A
similar amount (just slightly less) PRF is observed with {\it HST}'s ACS CCDs\footnotemark. Once
vibrations and PRF are minimized the images have 21-23 mas
resolutions.

\footnotetext{Section 5.6.1, ACS Instrument Handbook Cycle 19}

\begin{figure}
\includegraphics[angle=0,width=\columnwidth]{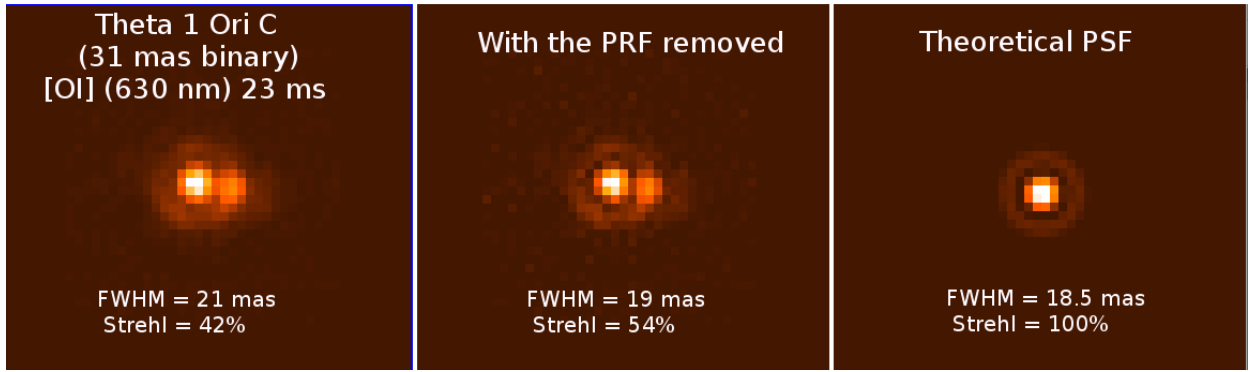}\caption{
An excellent short exposure single image of $\theta^{1}$ Ori C at [OI] (630 nm). On the left is the raw image with a resolution of $0.021\arcsec$, Strehl 42\%. Then the VisAO CCD's PRF is removed in the middle box and so the resolution is restored to the true value entering the CCD of $0.019\arcsec$. These are the highest resolution short exposure images ever obtained on any telescope to our knowledge.}  
\label{fig3b}
\end{figure}

Very short (23 msec) individual images were not effected as much by
the residual vibrations and achieved very high resolutions of 21 mas
(see Fig. \ref{fig3b}). These vibrations were found in commissioning to be mainly residual 60Hz vibrations not corrected by MagAO and are likely due to a few fans on the telescope that we could not turn off. However, once corrected for PRF these images are
diffraction-limited (FWHM=19 mas; Strehl=54\%; see Fig. \ref{fig3b}). We do not use Lucky imaging in this study, since the long exposure (60s) images in Fig. \ref{fig3} are much deeper (and almost as sharp) as those possible to obtain with Lucky in 60 s of telescope time.

\section{ASTROMETRY \& PHOTOMETRY}

\begin{figure}
\includegraphics[angle=0,width=\columnwidth]{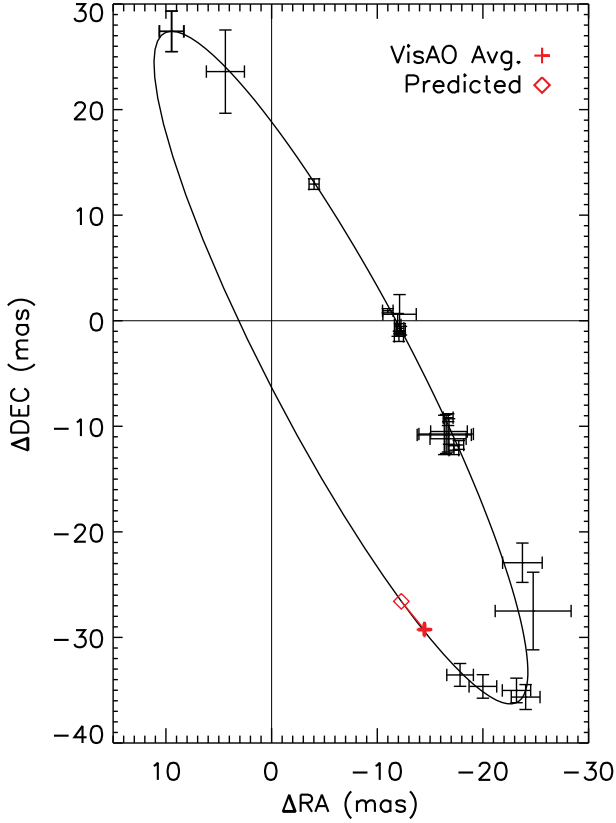}\caption{
PSF fitting photometry of the [OI], r', and $H_{\alpha}$ images in Fig. \ref{fig3} of $\theta^{1}$ Ori $C_1C_2$ with a 44\% Strehl theoretical PSF gave a good fit to both binary components in all cases. For Dec 12, 2012 UT we find separation is $32.64\pm0.48$ mas and PA is $206.31\pm0.17^{\circ}$. Here we plot this position (in red) against the interferometric orbit of \cite{kra09}. The agreement with the predicted position of $C_2$ w.r.t. $C_1$ is reasonable given the uncertainty of the orbital solution. However, more VisAO astrometry at this level of accuracy following this poorly sampled side ($20^{\circ}<$PA$<210^{\circ}$) of the orbit over the next few years would certaintly lead to a better orbital solution than is known today. }  
\label{orbit_C}
\end{figure}

All reduced (with SAA but not PRF corrected) 60s images of $\theta^{1}$ Ori $C_{1}C_2$ were analyzed with the
DAOPHOT PSF fitting task {\it allstar} \citep{ste87}. The $\pm0.48$ mas astrometric error of this very tight binary (where our $\sim 0.03 mas$ platescale errors can be ignored) was estimated by the standard deviation of the astrometry differences between the three filters ([OI], r', $H\alpha$\footnotemark) used. Our $\theta^{1}$ Ori $C_{1}C_2$ measurements of $32.64\pm0.48$ and $PA=206.31\pm0.17^{\circ}$ are compared to the interferometrically derived orbit in Fig. \ref{orbit_C}. We find reasonable agreement between the AO images and the interferometrically derived orbit of \cite{kra09}. For $\theta^{1}$ Ori $B_{1},B_2,B_3,B_4$ and $\theta^{1}$ Ori $A_{1},A_2$ 
the astrometry are summarized in Table \ref{tbl-1}. The
columns of Table \ref{tbl-1} are self explanatory.

\footnotetext{While calibrating the throughput of the $H\alpha$ filter in the second commissioning run we found a faint companion to the famous transition disk young star HD 142527. The position of this companion at 83 mas and $130^\circ$ PA was similar to a candidate companion found by aperture masking interferometry at the VLT by \cite{bil12} who measured $88\pm4$ mas, $133\pm3^\circ$. Hence, we report for the first time, that the existence of the close stellar companion HD 142527B is confirmed as real. Further details about this object are beyond the scope of this work but will be the focus of a future paper, Close et al. in prep.}

 In the  $\theta^{1}$ Ori $B$ group the PSF star used was the unsaturated $\theta^{1}$ Ori $B_{1}$ itself. Since all the members of
the $\theta^{1}$ Ori $B$ group are located within $1\arcsec$ of
$\theta^{1}$ Ori $B_{1}$ the PSF fit is particularly excellent there (there is no
detectable change in PSF morphology due to anisoplanatic effects
inside the $\theta^1$ Ori B group \citep{dio00}). Moreover, the residuals over the whole field were less than a few \% after PSF subtraction. This is not really surprising given the quality of the nights combined with the fact that no star was further than $\sim1\arcsec$ from the guide star. However, to minimize this affect, we only used the longer wavelength z' images reduced with SAA (taken on Dec 4, 2012) where anisoplanatic PSF effects were undetected. The relative positional accuracy is an excellent $\sim 0.2-1.4 $ mas in radial separation. The $\sim 0.2-1.4$ mas separation errors are the resultant of the platescale uncertainty added in quadrature with the measurement uncertainty ($FWHM/(S/N)$). The errors are somewhat worse in the PA direction ($0.6-5$ mas) due to a fixed $\pm0.3$ degree conservative estimate of our absolute rotator uncertainly.

\begin{figure}
\includegraphics[angle=0,width=0.4\columnwidth]{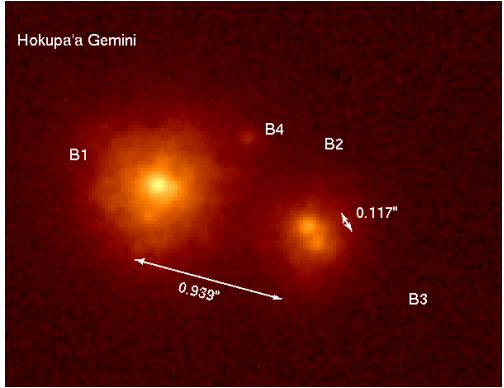}\caption{
The 8m Gemini/Hokupa'a images of the $\theta^{1}$ Ori B group in the $K^\prime$ band (09/19/01; from \cite{clo03c}). Resolution $0.085\arcsec$. Log scale. North is up and East is left. 
}
\label{fig4}
\end{figure}

\begin{figure}
\includegraphics[angle=0,width=0.4\columnwidth]{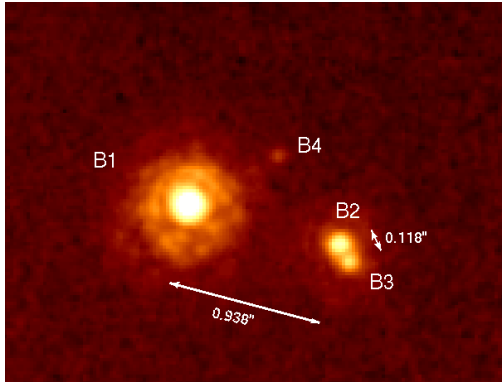}
\caption{
Detail of the $\theta^{1}$ Ori B group as imaged at $0.077\arcsec$ (Strehl $> 20\%$)
resolution (in the H band) with the MMT AO system (01/20/03) from \cite{clo03c}.}
\label{mmt_B}
\end{figure}

\begin{figure}
\includegraphics[angle=0,width=0.6\columnwidth]{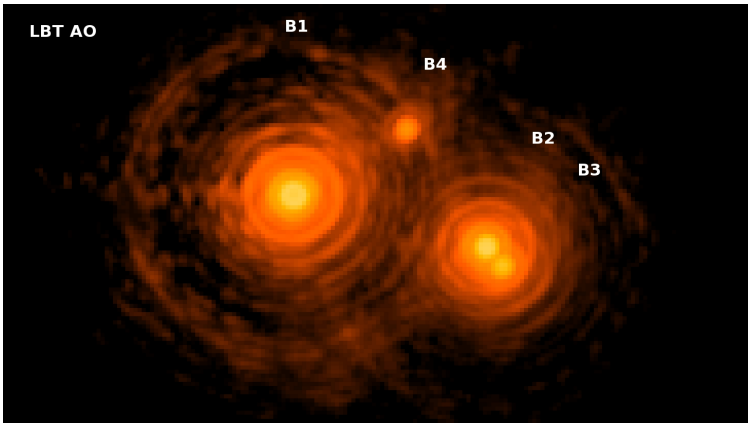}\caption{
The LBT AO $Br\gamma$ (2.16 $\mu m$) images of the $\theta^{1}$ Ori B group. Resolution $0.06\arcsec$. Logarithmic color scale. North is up and east is left. Strehl is $\sim75\%$ (from \cite{clo12b}). 
}
\label{lbt_B}
\end{figure}

\begin{figure}
\includegraphics[angle=0,width=0.6\columnwidth]{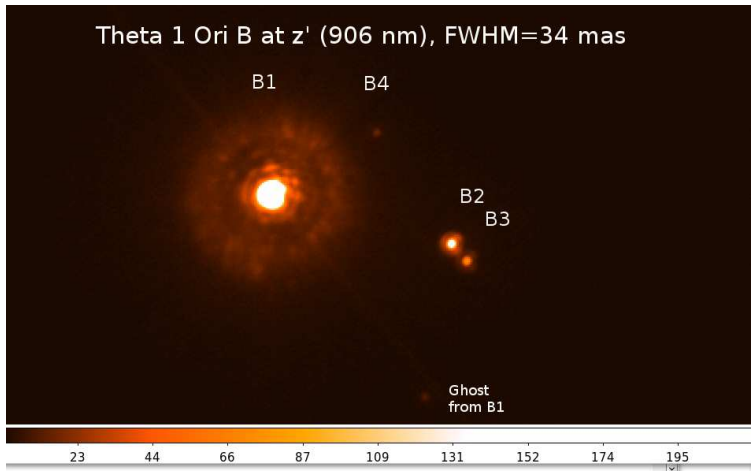}\caption{
The MagAO z' (0.91 $\mu m$) images of the $\theta^{1}$ Ori B group. Resolution $0.034\arcsec$. Linear color scale. North is up and east is left. Note that this image is $\sim2x$ sharper than that of the 8.4m LBT at $Br\gamma$ (2.16$\mu m$) (see fig \ref{lbt_B}). This is clear evidence that AO in the ``blue'' allows a smaller 6.5m telescope to outperform the resolution of an 8m in the NIR.
}
\label{mag_B}
\end{figure}

We can also compare our MagAO data to older (somewhat less accurate)
images of the Trapezium B stars from \cite{clo03c} who used AO images
from Gemini and the 6.5m MMT and speckle images from the
literature \citep{sch03}. Even though these individual observations
are of lower quality and Strehl than the MagAO ones (compare
Figs. \ref{fig4}, \ref{mmt_B}, and \ref{lbt_B} to that of MagAO in
Fig. \ref{mag_B}), the 15 years between these observations and those
of MagAO can highlight even very small orbital motions of bound
systems in the Trapezium. It also shows the very significant
improvement in high Strehl AO now possible with Pyramid wavefront
sensors and next generation adaptive secondary mirrors (ASMs).

\begin{figure}
\includegraphics[angle=270,width=0.5\columnwidth]{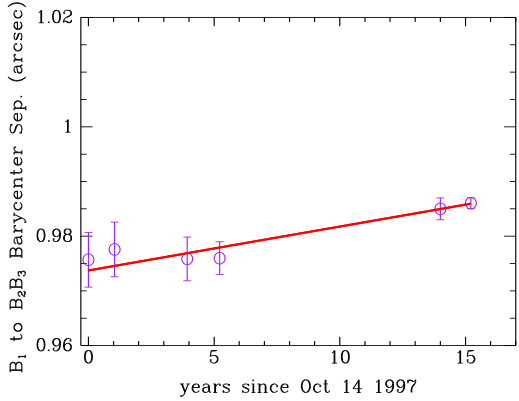}\caption{
The separation between $\theta^{1}$ Ori $B_1$ and the barycenter of the $B_2B_3$ binary. Note how over
15 years of observation there has been a small, yet significant, relative
proper motion observed (0.80$\pm0.18 $ mas/yr; which is
a significant correlation at the 97.4\% level). The first 2 data points are speckle
observations from the 6-m SAO telescope \citep{wei99}, the next point
is from Gemini/Hokupa'a observations \cite{clo03c} followed by MMT AO observations \cite{clo03c}, and then LBT \cite{clo12b}, and the last from the MagAO system (this study). }
\label{fig6}
\end{figure}

\begin{figure}
\includegraphics[angle=270,width=0.5\columnwidth]{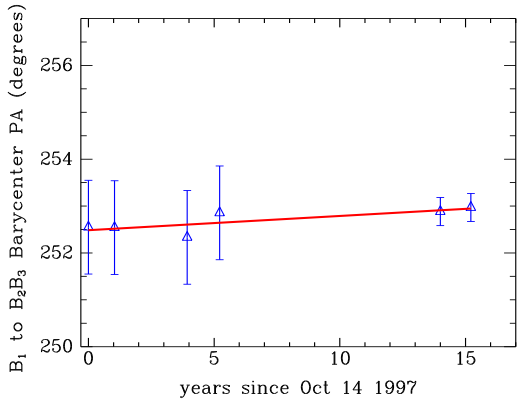}\caption{
The position angle between $\theta^{1}$ Ori $B_1$ and the barycenter of the $B_2B_3$ binary. Note how
over 15 years of observation there has been little relative
PA motion observed (0.030$\pm0.044^\circ$/yr which is just significant at the 88\% level). The epochs of the data are the same as in Fig. \ref{fig6}.}
\label{fig7}
\end{figure}

\begin{figure}
\includegraphics[angle=270,width=0.6\columnwidth]{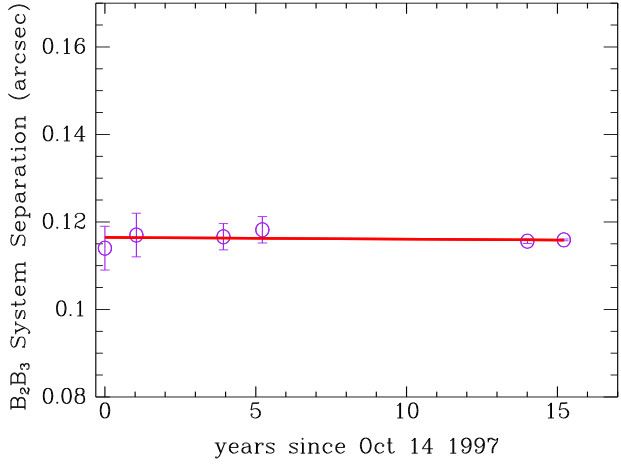}\caption{
The separation between the $\theta^{1}$ Ori $B_2$ and $B_3$ components. Note the surprising lack
of any significant relative motion ($-0.04\pm0.14$ mas/yr). The rms scatter from a constant value is only $0.14$ mas/yr. There appears to be very little change in the separation of the $B_2B_3$ system. The epochs of the data are the same as in Fig. \ref{fig6}. }
\label{fig8}
\end{figure}

\begin{figure}
\includegraphics[angle=270,width=0.6\columnwidth]{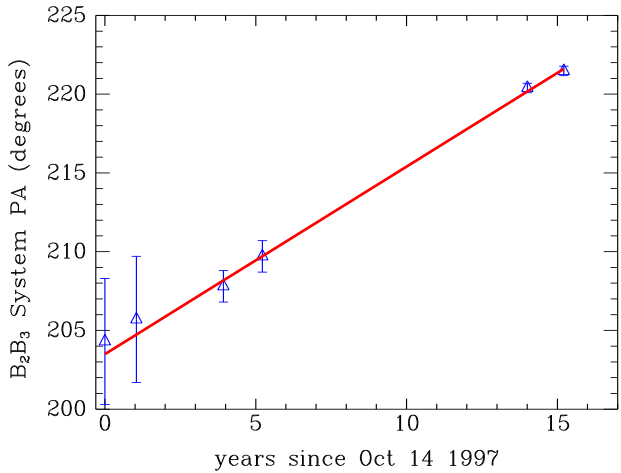}\caption{
The position angle of $\theta^{1}$ Ori $B_2$ and $B_3$. Here we
observe, clearly, real orbital arc of motion where $B_3$ moving
counter-clockwise (at 1.19$\pm0.06^\circ$/yr; a correlation significant at the
99.9\% level) around $B_2$. The epochs of the data are the same as in Fig. \ref{fig6}. }
\label{fig9}
\end{figure}

\begin{figure}
\includegraphics[angle=270,width=0.6\columnwidth]{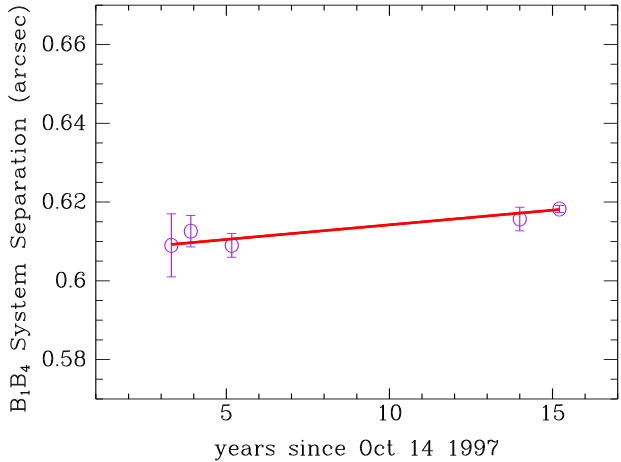}\caption{
The separation between $\theta^{1}$ Ori $B_1$ and $B_4$. Note how over
13 years of observation there is little change in the separation
(0.72$\pm0.23$ mas/yr; correlation 95\%). The first data point is an speckle observation from the
6-m SAO telescope \citep{sch03}, then Gemini/Hokupa'a observation \cite{clo03c}, the next data point is from the MMT AO observation \cite{clo03c}, the next from the LBT \cite{clo12b}, and the last is from MagAO. }
\label{figB4_sep}
\end{figure}

\begin{figure}
\includegraphics[angle=270,width=0.6\columnwidth]{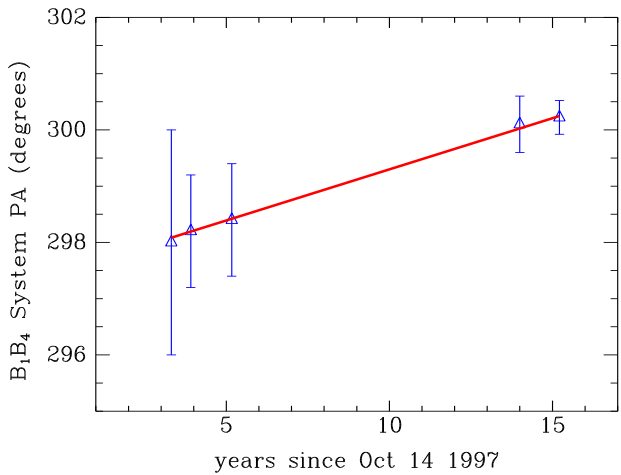}\caption{
The position angle between $\theta^{1}$ Ori $B_1$ and $B_4$. Note how
over 13 years of observation there has been only now clear significant relative
proper motion observed (0.181$\pm0.067^\circ$/yr; correlation 99.8\%). The sources of the data is the same as in Fig. \ref{figB4_sep}. }
\label{figB4_pa}
\end{figure}

A test to see how accurate our astrometry is over the last 15 years is to look at the scatter from a linear trend of the $\theta^{1}$ Ori $B$ group's motions. A comparison of our highly accurate positions with the historical positions from the literature is summarized in Table
\ref{tbl-1}. Linear (weighted by astrometric error) fits to the
data in Table \ref{tbl-1} (Figures \ref{fig6} to \ref{figB4_pa})
yield the velocities shown in Table \ref{tbl-1}. The overall error in
the relative proper motions is now an impressive $\la 0.2$ mas/yr in
proper motion ($\la 0.4$ km/s) a factor of 2 improvement in accuracy when the VisAO positions are added into these calculations, compared to the last published values from \cite{clo12b}.

\section{ANALYSIS \& DISCUSSION}

With these accuracies it is now possible to determine whether these
stars in the $\theta^{1}$ Ori $B$ group are bound together, or merely
chance projections in this very crowded region. We adopt the masses of each star from the \cite{sie97,ber96} tracks
fit by \cite{wei99} where we find masses of: $B_1\sim7M_{\sun} $;
$B_2\sim3M_{\sun} $; $B_3\sim2.5M_{\sun} $; $B_4\sim0.2M_{\sun} $;
$B_5\sim 7 M_{\sun}$; $A_1\sim20M_{\sun} $; $A_2\sim 4M_{\sun} $; and
$A_3\sim 2.6 M_{\sun}$. Based on these masses (which are similar to those adopted by \cite{sch03}) we can comment on
whether the observed motions are less than the escape velocities
expected for simple face-on circular orbits.

Our combination of high spatial resolution and high signal to noise
shows that there is very little significant motion in the $B_1B_2$ system over
the last 15 years (as we might expect since the rotation of $B_2$ about the barycenter of the $B_2B_3$ system appears to be just canceling the motion of $B_2$ w.r.t. $B_1$). But of course it is really the {\it barycenter} of the tight $B_2B_3$ binary that would be in orbit around $B_1$. Hence the barycenter would have to show steady orbital motion if bound to $B_1$. Since $B_2$ is only 20\% more massive than $B_3$ this means the $B_2B_3$ barycenter is currently 52 mas at PA $221.5^{\circ}$ from the center of $B_2$. In figures \ref{fig6} and \ref{fig7} we see that there is a small, yet significant, motion of the barycenter of $B_2B_3$ w.r.t. $B_1$ of some $0.80\pm0.18$ mas/yr ($1.6\pm0.3$km/s) and in PA by $0.030\pm0.044^{\circ}$/yr ($1.0\pm1.0$ km/s). Hence the motion of $B_2B_3$ is currently about $1.9\pm0.6$ km/s in the direction of PA $\sim305^{\circ}$ (moving towards the WNW direction from $B_1$). This is the first time this motion has been detected. At this time it is not yet possible to prove this is true orbital motion, but given how close $B_2B_3$ is to $B_1$ this is likely orbital motion.

We have, of course, observed clear orbital
motion (at $4.7\pm0.2$ km/s) in the very tight $\theta^{1}$ Ori $B_{2}B_{3}$ system in almost pure PA
(see Figure \ref{fig9}). In fact, now that we have observed over $20^{\circ}$ of PA rotation (with no significant change in separation), we have clear evidence of an ``arc'' of curvature of the system. The motion of the $B_2B_3$ binary is roughly consistent with a face-on, circular orbit (orbiting in the counterclockwise direction). A mildly elliptical orbit is also quite plausible given the very small amount of orbital phase observed to date.  

Also we see linear orbital motion of $7.0\pm0.5$ km/s in the $\theta^{1}$ Ori $A_{1}A_{2}$
system (see Table \ref{tbl-1}). This is consistent with the motion seen by \cite{gil13}. We know this is likely orbital motion since it is higher than the motion of unrelated stars in the cluster, due to their very close separation of just $0.19\arcsec$.



\subsection{Is the $\theta^1$ Ori $B_2B_3$ System Physical?}

The relative velocity in the $\theta^{1}$ Ori $B_{2}B_{3}$ system (in
the plane of the sky) is now more accurate by $\sim10\times$ compared to that of \cite{clo03c} and by $\sim2\times$ compared to \cite{clo12b}. Our new velocity of $4.7\pm0.2$ km/s is consistent, but with much lower errors, with the $\sim 4.2\pm2.1$ km/s of \cite{clo03c} (this velocity is in the azimuthal direction; see Figure \ref{fig9}). This is a reasonable
$V_{tan}$ since an orbital velocity of $\sim 6.7$ km/s is expected
from a face-on circular orbit from a $\sim5.5 M_{\sun}$ binary system
like $\theta^{1}$ Ori $B_{2}B_{3}$ with a 49 AU projected
separation (implying an orbital period of order $\sim200$ yr). This theoretical value of $\sim200$ yr is close to the $302\pm16$ yr orbit that comes from assuming the current measured angular (PA) velocity stays constant. It is worth noting that this velocity is also greater than
the $\sim 3$ km/s \cite{hil98} velocity dispersion of the cluster.


Our observed velocity of $1.19\pm0.06^{\circ}$/yr is $\sim10\times$ more accurate than that of \cite{clo03c}. This primarily azimuthal motion {\it strongly} suggests a curving orbital arc of $B_3$
orbiting $B_2$ counterclockwise.




\subsection{Is the $\theta^1$ Ori B Group Stable Long Term?}

The barycenter of $B_2B_3$ is moving at a low  $1.9\pm0.6$ km/s in the
plane of the sky w.r.t. to $B_1$ (itself a very tight pair with $B_5$) where the escape 
velocity $V_{\mathrm esc} \sim 6$\,km/s for this massive system ($\sim20 M_{\odot}$). Hence these two
pairs are likely gravitationally bound together. This is the first effort that has measured this small barycenter velocity definitively. Hence, we can say that these two pairs currently form a rare bound ``mini-cluster'' of young massive stars. 

\subsubsection{Is the Orbit of $\theta^1$ Ori $B_4$ Stable?}

The two AO measurements of \cite{clo03c} (and the one speckle detection of \cite{sch03}) did not detect a
significant velocity of $B_4$ w.r.t. $B_1$: $2\pm11$ km/s. However, our much better data and timeline between the LBT epoch and the excellent MagAO observations has shed some light on the
question of B4 orbiting B1. As is clear from
Figures \ref{figB4_sep} \& \ref{figB4_pa}, there appears to be a
real velocity of $4.1\pm1.2$ km/s detected. This is greater than the
random velocity of the cluster yet below the escape velocity of $\sim
6$ km/s, this points towards $B_4$ being also gravitationally bound member
of the $\theta^1$ Ori B group. Again we are observing almost pure motion in PA ($0.181^{\circ}$/yr counterclockwise). Assuming a simple face-on orbit we would expect a very rough period of $\sim2000\pm700$ years for $B_4$ to orbit $B_1$ given an average angular velocity of $0.181^{\circ}$/yr. 

\subsection{A Possible Model for the Orbits of the $\theta^1$ Ori $B$ Group}

It is tempting to define a circular orbit baseline model of the
$\theta^1$ Ori $B$ system with the center as the very tightly bound
6.47 day massive ($\sim14 M_{sun}$) 0.13 AU spectroscopic binary
$B_1B_5$ which we cannot spatially resolve. Around this center is the
low mass $\sim0.2 M_{sun}$ $B_4$ some $\sim254$ AU away which orbits every
$\sim2000\pm700$ years in a roughly face-on circular orbit. Then,
further out, the tight 49 AU binary $B_2B_3$ rotates every $302\pm16$
years around its barycenter with roughly a face-on circular
orbit. However, the co-planar geometry is broken by the orbit of this
barycenter around $B_1$. It appears that the $B_2B_3$ barycenter is
moving in a bound orbit to WNW (PA $\sim305^{\circ}$). This motion
cannot be in a simple face-on circular orbit, and so must be (if close
to circular) inclined by about $\sim30^{\circ}$, but many other
elliptical orbits are also possible. We simply do not have enough of
time baseline to understand the fine details of this orbit today. If
we simply assume it is a inclined circular orbit then it has roughly a
$\sim820AU$ (deprojected) separation from $B_1$ and a period of some
$\sim 11,000$ years.

\begin{figure}
\includegraphics[angle=0,width=0.6\columnwidth]{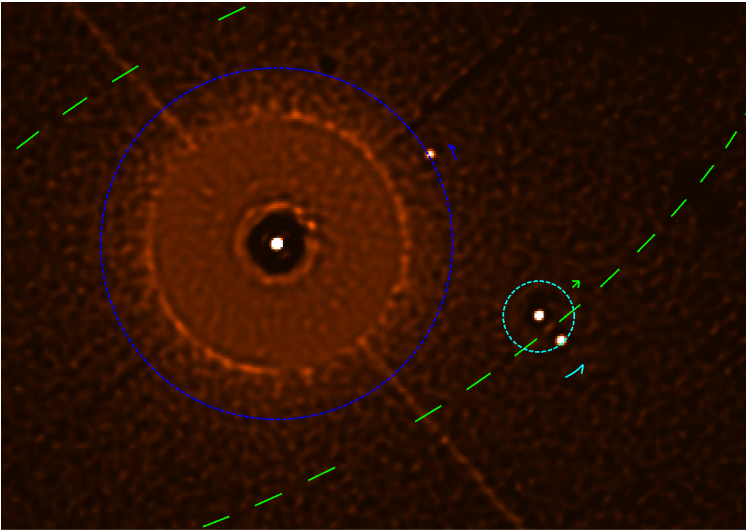}\caption{
A possible model of the motions of the $\theta^{1}$ Ori $B$ group. Here we show a {\it lucy} deconvolved image of the z' image. We make an assumption that all the orbits are circular and plot possible orbital solutions for each component's orbit about $B_1$ based on this rough assumption. We also plot the actual observed orbital ``arcs'' so far imaged over the last 15 years for the system. Clearly the orbits are still undefined, but this plot gives some insight into the nature of the system. }
\label{model1}
\end{figure}

\begin{figure}
\includegraphics[angle=0,width=0.6\columnwidth]{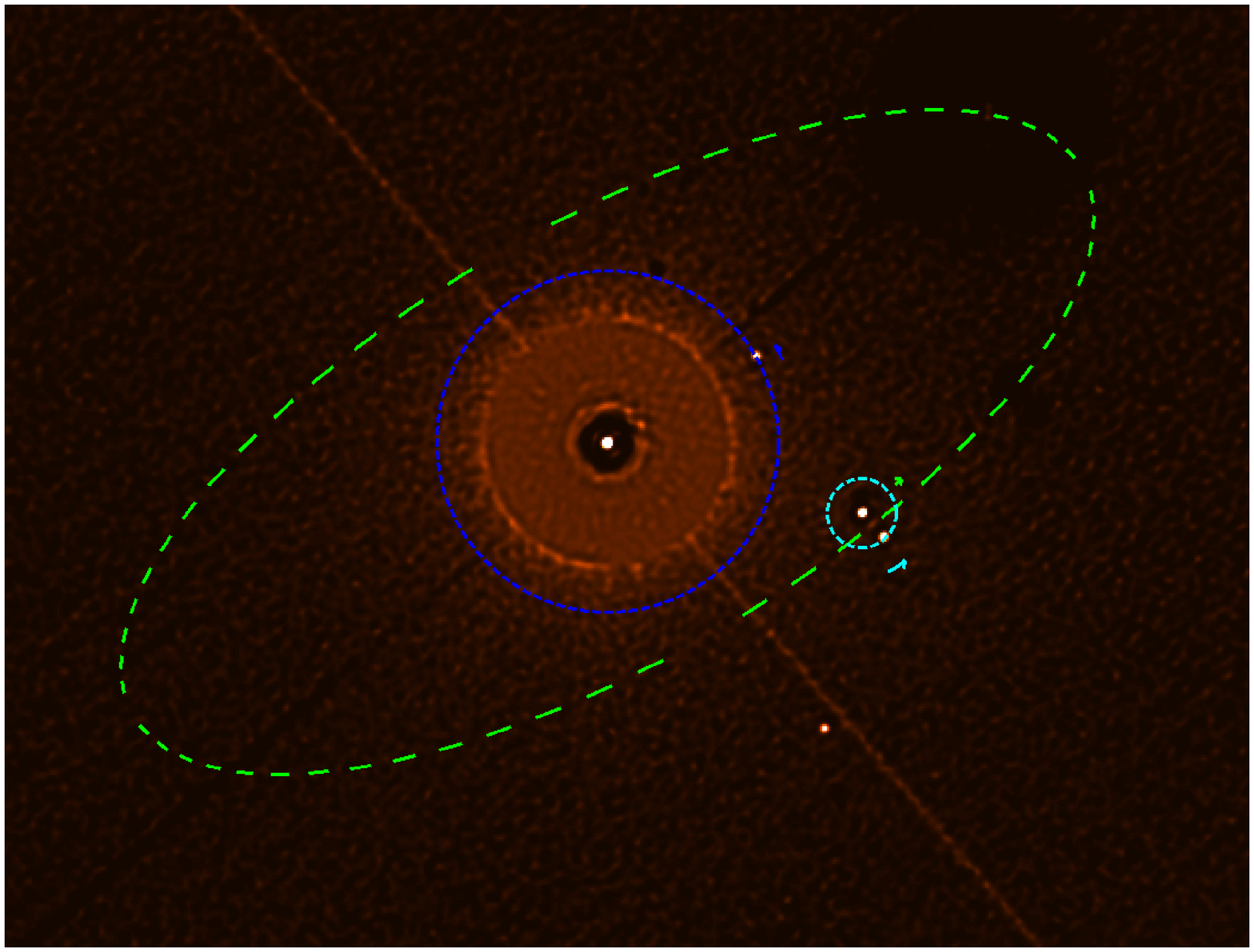}\caption{
A zoom out of Fig. \ref{model1} to show the full inclined orbit of the $B_2B_3$ barycenter around $B_1$. }
\label{model2}
\end{figure}

See Figs \ref{model1} and \ref{model2} for illustrations of what these
obits would look like if they are all close to circular. It is
interesting to note, that once the true (deprojected) separation of $B_2B_3$ is
considered, the group seems more hierarchical than reported
in \cite{clo12b}. For example, the ratio of the 3 main periods are
$P_{23} : P_{1/4} : P_{1/23}$ = 1 : 7 : 36 so that there is almost an
order of magnitude separating each period. This large spread of
orbital periods will lend some stability to this ``mini-cluster''. On
the other hand, $B_4$'s very low mass, its intermediate period, and
its location w.r.t. to the other four groups members makes it highly
unlikely that $B_4$ is on a long-term stable orbit within the
group. It is very likely that an interaction between the much more
massive $B_2B_3$ and $B_4$ will eject $B_4$ in the future --leading to
a slightly more tightly bound ``mini-cluster'' without $B_4$. As we
will discuss in the next section, even the much more massive $B_3$ may
not even be stable in the long-term.

\subsubsection{Is the orbit of $B_3$ around $B_2$ and of $B_5$ around
$B_1$ stable in the long-term?}

\cite{clo12b} noted that the distance $D_{B_1B_5} \sim 3 \times 10^{-4} \times D_{B_1B_5 B_2B_3}$ and thus
the very tight (0.13 AU) tight $B_1B_5$ system is, of course, very stable. More interesting is the case of $B_2B_3$. Their de-projected distance is not very small compared to their
projected distance (D) from the $B_1B_5$ pair:$ D_{B_2B_3} \sim 0.06 \times D_{B_1B_5 B_2B_3}$. Thus the stability of the $B_2B_3$ orbit needs a more
detailed analysis since it is possible that $B_3$ may be ejected in the future.


The orbital period of the two binaries w.r.t. each other is 
$P_{1/23} \sim 11000\,$yrs, while the orbital period of $B_3$ w.r.t 
$B_2$ amounts to $P_{2/3} \sim 300\,$yrs. For the calculation of both 
periods, we have assumed the masses as given above, and circular orbits 
in the plane of the sky (except for $B_1B_3$ which is inclined at $\sim30^{\circ}$). This leads to a period ratio $X = 
P_{1/23}/P_{2/3} \sim 36$. Eggelton \& Kiseleva's stability 
criterion requires $X \ge  X_{\mathrm crit} = 10.08$ for the masses 
in the B group. This means that within the accuracy limits of our 
investigation, the binary $B_2B_3$ is likely stable (different from the marginal stability found in \cite{clo12b}). The stability criterion depends also on the orbits' 
eccentricities. However, mild eccentricities of the order
of $e \sim 0.1$ (as can be expected to develop in hierarchical triple 
systems; see, e.g., Georgakarakos 2002), can make the B group unstable. However, the $\theta^1$ Ori B system seems to be a good example of a highly
dynamic star formation "mini-cluster" which might, in the future,
eject the lowest-mass member(s) through dynamical decay
\citep{dur01}.

\section{Conclusions}

In this study we utilized the new high-order (585 actuator) Magellan Adaptive Optics system (MagAO)
to obtain very high-resolution science in the visible with MagAO's
VisAO CCD camera. In the median seeing conditions of Magellan ($0.5-0.7\arcsec$)
we found that MagAO delivers individual short exposure images as good as 19 mas
optical resolution. Due to residual 60Hz vibrations, long exposure (60s) r' (0.63$\mu m$) images are slightly
coarser at FWHM=23-29 mas (Strehl $\sim28\%$) with bright ($R<9$ mag) guide stars. These are the highest resolution filled-aperture images
published to date.  Images of the young ($\sim 1$ Myr) Orion Trapezium
$\theta^{1}$ Ori A, B, and C cluster members were obtained with the
VisAO camera. In particular, the 32 mas binary $\theta^{1}$ Ori $C_1C_2$ was
easily resolved in non-interferometric images for the first time. Relative
positions of the bright trapezium binary stars were measured with 0.6-5.0 mas accuracy. We now are
sensitive to relative proper motions of just $\sim0.2$ mas/yr
($\sim0.4$ km/s at 414 pc) -- this is a $\sim2-10\times$ improvement in
velocity accuracy compared to previous efforts.
We now
detect clear orbital motions of $\theta^{1}$ Ori $B_{2}B_{3}$ and
$A_1A_2$ of $4.7\pm0.2$ km/s and $7.1\pm0.5$ km/s, respectively. 
For the first time, we see clear motion of the barycenter of
$\theta^{1}$ Ori $B_{2}B_{3}$ in about $\theta^{1}$ Ori
$B_{1}$. All five members of the $\theta^{1}$ Ori $B$ system appear
likely a gravitationally bound ``mini-cluster'', but we find that not
all the orbits can be both circular and co-planar. The very lowest mass
member of the $\theta^{1}$ Ori $B$ system ($B_{4}$; mass $\sim 0.2
M_{\sun}$) has a very clearly detected motion (at $4.1\pm1.3$ km/s;
correlation=99.9\%) w.r.t $B_1$. Previous work has suggested that $B_4$
and $B_3$ are both on long-term unstable orbits and will be ejected from
this ``mini-cluster''.  However, our new ``baseline'' model of the
$\theta^{1}$ Ori $B$ system suggests a more hierarchical system than
previously thought, and so the ejection of $B_4$ may not occur for
many orbits, and $B_3$ may be stable against ejection long-term. This
``ejection'' process of the lowest mass member of a ``mini-cluster'' could play a
major role in the formation of low mass stars and brown dwarfs.

\section {FUTURE OBSERVATIONS}

Future observations are required to see, if indeed, these stars continue
to follow orbital arcs around each other proving that they are interacting
with one another. In addition, future observations of the $\theta^1$
Ori $B_4$ positions would help deduce if it is on a marginally stable orbit given its somewhat ``non-hierarchical'' location in the B group. 

Future observations should also try to determine the radial velocities
of these stars. Once radial velocities are known one can calculate
the full space velocities of these stars. Such observations will
require both very high spatial and spectral resolutions. This might be
possible with such instruments like the AO fed ARIES echelle
instrument at the MMT.

\acknowledgements

We thank the whole Magellan Staff for making this wonderful telescope
possible for use with our AO system. We would especially like to thank Povilas Palunas (for help
over the entire MagAO commissioning run). Juan Gallardo (and his professional team), Patricio
Jones, Emilio Cerda, Felipe Sanchez, Gabriel Martin, Maurico
Navarrete, Jorge Bravo and the whole team of technical experts helped do
many exacting tasks in a very professional manner. Glenn Eychaner,
David Osip and Frank Perez all gave expert support which was
fantastic. Also thanks to Victor, Maurico, and Hugo for running the
telescope so well. It is a privilege to be able to commission an AO
system on such a fine telescope and site. Thanks from the whole MagAO
team. And thanks to Miguel Roth, Francisco Figueroa, Roberto Bermudez,
Sergio Veliz and, of course, Mark Philips for making this
all happen -- and very smoothly -- despite the large AO team that was
needed at the mountain and all the headaches and extra work it created
for the LCO staff. We also thank the teams at Steward Observatory Mirror Lab/CAAO (University of Arizona), Microgate (Italy) and ADS (Italy) for building such a great ASM. The MagAO ASM was developed with support from
the NSF MRI program. The MagAO PWFS was developed with help from the NSF TSIP program and the Magellan partners. The Active Optics guider was developed by Carnegie Observatories with custom optics from the MagAO team. The VisAO camera and commissioning was supported with
help from the NSF ATI program. LMC's research was supported by NSF AAG
and NASA Origins of Solar Systems grants. JRM is grateful for the
generous support of the Phoenix ARCS Foundation. KM was supported
under contract with the California Institute of Technology (Caltech)
funded by NASA through the Sagan Fellowship Program.





\clearpage

\clearpage
\begin{deluxetable}{lllllllll}
\tabletypesize{\scriptsize}
\tablecaption{High Resolution Observations of the $\theta^{1}$ Ori B \& A groups\label{tbl-1}}
\tablewidth{0pt}
\tablehead{
\colhead{System} &
\colhead{$\Delta H$} &
\colhead{$\Delta K^{\prime}$} &
\colhead{Separation} &
\colhead{Sep. Vel.} &
\colhead{PA} &
\colhead{PA Vel.} &
\colhead{Telescope} &
\colhead{epoch}\\
\colhead{name} &
\colhead{(mag)} &
\colhead{(mag)} &
\colhead{($\arcsec $)} &
\colhead{(Sep. mas/yr)} &
\colhead{($^{\circ}$)} &
\colhead{($^{\circ}$/yr)} &
\colhead{} &
\colhead{(m/d/y)}\\
}
\startdata

$B_{1}B_{2}$&$2.30\pm0.15$&&$0.942\pm0.020\arcsec$&&$254.9\pm1.0$&&SAO\tablenotemark{a}&10/14/97\\
&&$1.31\pm0.10$\tablenotemark{b}&$0.942\pm0.020\arcsec$&&$254.4\pm1.0$&&SAO\tablenotemark{a}&11/03/98\\
&&$2.07\pm0.05$&$0.9388\pm0.0040\arcsec$&&$254.6\pm1.0$&&GEMINI&09/19/01\\
&$2.24\pm0.05$&&$0.9375\pm0.0030\arcsec$&&$255.1\pm1.0$&&MMT&01/20/03\\
&&&$0.9411\pm0.0023\arcsec$&&$254.55\pm0.3$&&LBT&10/16/11\\
&&&$0.9415\pm0.0014\arcsec$&&$254.64\pm0.3$&&MagAO&12/04/12\\
&&&&&&&&\\
&&& with MagAO=&0.31$\pm0.25$&&-0.009$\pm0.043$&&\\
&&&corr.=& 89\%; no vel. & corr.=& 33\%; no vel. &&\\
&&&& detected\tablenotemark{c} && detected\tablenotemark{c} &&\\
\hline
&&&&&&&&\\
$B_{2}B_{3}$&$1.00\pm0.11$&&$0.114\pm0.05\arcsec$&&$204.3\pm4.0$&&SAO\tablenotemark{a}&10/14/97\\
&&$1.24\pm0.20$&$0.117\pm0.005\arcsec$&&$205.7\pm4.0$&&SAO\tablenotemark{a}&11/03/98\\
&&$1.04\pm0.05$&$0.1166\pm0.0040\arcsec$&&$207.8\pm1.0$&&GEMINI&09/19/01\\
&$0.85\pm0.05$&&$0.1182\pm0.0030\arcsec$&&$209.7\pm1.0$&&MMT&01/20/03\\
&&&$0.1156\pm0.0005\arcsec$&&$220.39\pm0.3$&&LBT&10/16/11\\
&&&$0.1160\pm0.0002\arcsec$&&$221.50\pm0.3$&&MagAO&12/04/12\\
&&&&&&&&\\
&&&with MagAO=&-0.04$\pm0.14$&&1.19$\pm0.06$&&\\
&&&corr.=&24\%; no vel.&&{\bf 4.7$\pm$0.2 km/s}&&\\
&&&& detected && {\bf corr.=99.9\%} &&\\
\hline
&&&&&&&&\\
$B_{1}B_{4}$&&$5.05\pm0.8$&$0.609\pm0.008\arcsec$&&$298.0\pm2.0$&&SAO\tablenotemark{d}&02/07/01\\
&&$5.01\pm0.10$&$0.6126\pm0.0040\arcsec$&&$298.2\pm1.0$&&GEMINI&09/19/01\\
&$4.98\pm0.10$&&$0.6090\pm0.0050\arcsec$&&$298.4\pm1.0$&&MMT&01/20/03\\
&&&$0.6157\pm0.003\arcsec$&&$300.1\pm0.5$&&LBT&10/16/11\\
&&&$0.6182\pm0.0009\arcsec$&&$300.23\pm0.3$&&MagAO&12/04/12\\
&&&&&&&&\\
&&&with MagAO=&0.72$\pm0.23$&&0.181$\pm0.067$&&\\
&&&&{\bf 1.4$\pm$0.5 km/s}&&{\bf 3.83$\pm$1.27 km/s}&&\\
&&&&{\bf corr.=95\%} && {\bf corr.=99.9\%} &&\\
\hline
&&&&&&&&\\
$A_{1}A_{2}$&$1.51\pm0.15$&$1.38\pm0.10$&$0.208\pm0.030\arcsec$&&$343.5\pm5.0$&&Calar Alto\tablenotemark{e}&11/15/94\\
&&$1.51\pm0.05$&$0.2215\pm0.005\arcsec$&&$353.8\pm2.0$&&SAO\tablenotemark{a}&11/03/98\\
&&$1.62\pm0.05$&$0.2051\pm0.0030\arcsec$&&$356.9\pm1.0$&&GEMINI&09/19/01\\
&&&$0.1931\pm0.0005\arcsec$&&$366.5\pm0.3$&&LBT&10/16/11\\ 
&&&$0.1881\pm0.0016\arcsec$&&$367.6\pm0.3$&&MagAO\tablenotemark{f}&12/08/12\\ 
&&&&&&&&\\
&&&with MagAO=&-1.6$\pm0.2$&&0.98$\pm0.07$&&\\
&&&&{\bf -3.2$\pm$0.3 km/s}&&{\bf 6.3$\pm$0.4 km/s}&&\\
&&&&{\bf corr.= 92.9\%}&&{\bf corr.= 99.4\%}&&\\
\enddata
\tablenotetext{a}{speckle observations of \cite{wei99}.}
\tablenotetext{b}{these low $\Delta K$ values are possibly due to $\theta^{1}$ Ori $B_{1}$ being in eclipse during the 11/03/98 observations of \cite{wei99}. }
\tablenotetext{c}{Note there is velocity detected from $B_1$ w.r.t. the barycenter of the $B_2B_3$ binary see fig \ref{fig6} and fig \ref{fig7}.}
\tablenotetext{d}{speckle observations of \cite{sch03}.}
\tablenotetext{e}{speckle observations of \cite{pet98}.}
\tablenotetext{f}{$A_1A_2$ Data from Ks image from the MagAO/Clio2 NIR camera (Morzinski et al. 2013 in prep).} 
\end{deluxetable}

\clearpage

\clearpage

\clearpage

\clearpage

\clearpage

\clearpage

\clearpage

\clearpage

\clearpage




\begin{thebibliography}{}

\bibitem[Abt, Wang, \& Cardona(1991)]{abt91}Abt H.A., Wang R., Cardona O., 1991, \apj, 367, 155
\bibitem[Barnec et al.(2012)]{bar12}Baranec, C.; Riddle, R.; Ramaprakash, A. N.; Law, N.; Tendulkar, S.; Kulkarni, S.; Dekany, R.; Bui, K.; Davis, J.; Burse, M.; Das, H.; Hildebrandt, S.; Punnadi, S.; Smith, R. Proc SPIE Volume 8447, id. 844704-844704-11 
\bibitem[Bate et al.(2002)]{bat02}Bate, M.R., Bonnell, I.A., Bromm, V. 2002, \mnras, 332, L65
\bibitem[Bate et al.(2003)]{bat03}Bate, M.R., Bonnell, I.A., Bromm, V. 2003, \mnras, 277, 362
\bibitem[Bate et al.(2009)]{bat09}Bate, M.R. 2009, \mnras, 392, 590
\bibitem[Bate et al.(2011)]{bat11}Bate, M.R. 2011, \mnras, in press
\bibitem[Bazzon et al.(2012)]{baz12} Bazzon, Andreas; Gisler, Daniel; Roelfsema, Ronald; Schmid, Hans M.; Pragt, Johan; Elswijk, Eddy; de Haan, Menno; Downing, Mark; Salasnich, Bernardo; Pavlov, Alexey; Beuzit, Jean-Luc; Dohlen, Kjetil; Mouillet, David; Wildi, François (2012) Proceedings of the SPIE, Volume 8446, id. 844693-844693-14
\bibitem[Bernasconi \& Maeder(1996)]{ber96}Bernasconi P.A., \& Maeder A. 1996, A\&A, 307, 829
\bibitem[Biller et al.(2012)]{bil12}Biller, B., et al. 2012, \apj, 753, L38
\bibitem[Brusa et al.(2003a)]{bru03a}Brusa, G., et al. 2003a, Proc. SPIE 4839, 691.
\bibitem[Brusa et al.(2003b)]{bru03b}Brusa, G., et al. 2003b, Proc. SPIE 5169, 26
\bibitem[Burgasser et al.(2003)]{bur03}Burgasser, A. et al. 2003, \apj, 586, 512
\bibitem[Burgasser et al.(2007)]{bur07}Burgasser, A., Reid, L.N., Siegler, N., Close, L., Allen, P., Lowerance, P., Gizis, J. (2007),Protostars and Planets V, B. Reipurth, D. Jewitt, and K. Keil (eds.), University of Arizona Press, Tucson, 951 pp., p.427-441 
\bibitem[Burrows et al.(2000)]{bur00}Burrows, A., Hubbard, W. B., Lunine, J. I., Marley, M. S., Saumon, D. 
2000, Protostars and Planets IV (Tucson: University of Arizona Press, eds Mannings, V., Boss, A.P., Russell, S. S.), 
p. 1339
\bibitem[Bouy et al.(2003)]{bou03}Bouy, H., Brandner W., Mart\'\i n, E., Delfosse, X., Allard, F., \& Basri, G. 2003, \aj, 126, 1526
\bibitem[Bossi et al.(1989)]{bos89}Bossi M., Gaspani A., Scardia M., Tadini M., 1989, A\&A, 222, 117
\bibitem[Chabrier et al.(2000)]{cha00}Chabrier, G., Baraffe, I., Allard, F., \& Hauschildt, P. 
2000, \apj, 542, 464 
\bibitem[Close et al.(1998)]{clo98a}Close, L.M., Roddier, F.J., Roddier, C.A., Graves, J.E., Northcott, M.J., Potter, D. 
1998, Proc. SPIE Vol. 3353, p. 406-416. Adaptive Optical System Technologies, D. Bonaccini, R.K. Tyson, Eds
\bibitem[Close(2000)]{clo00}Close, L. M. 2000, Proc. SPIE Vol. 4007, p758-772.
 Adaptive Optical Systems Technology, P.L. Wizinowich, Ed.
\bibitem[Close et al.(2002a)]{clo02a}Close, L.M. et. al. 2002a, \apj, 566, 1095.
\bibitem[Close et al.(2002b)]{clo02b}Close, L.M. et. al. 2002b, \apj, 567, L53.
\bibitem[Close et al.(2003a)]{clo03a}Close, L.M., Siegler, N., Freed, M., Biller, B. 2003a \apj, 587, 407
\bibitem[Close et al.(2003b)]{clo03b}Close, L.M. et al. 2003b \apj, 598, 35.
\bibitem[Close et al.(2003c)]{clo03c}Close, L.M. et al. 2003c \apj, 599, 537.
\bibitem[Close et al.(2007)]{clo07}Close, L.M. et al. 2007 \apj, 665, 736.
\bibitem[Close et al.(2012a)]{clo12a}Close, L.M. et al. 2012a Proc. SPIE, Volume 8447, id. 84470X-84470X-16
\bibitem[Close et al.(2012b)]{clo12b}Close, L.M. et al. 2012b \apj, 749, article id. 180.
\bibitem[Dekany et al.(2013)]{dek13}Dekany, R. et al. Third International Conference on Adaptive Optics for Extremely Large Telescopes, in prep. 
\bibitem[Diolaiti et al.(2000)]{dio00}Diolaiti, E., Bendinelli, O. Bonaccini, D.; Close, L Currie, D. Parmeggiani, G. 2000, A\&AS 147, 335
\bibitem[Durisen, Sterzik, \& Pickett (2001)]{dur01}Durisen, R.H., Sterzik, M.F., \& Pickett, B.K. 2001, A\&A, 371, 952
\bibitem[Duquennoy \& Mayor(1991)]{duq91}Duquennoy, A., Mayor, M. 1991, \aap, 248, 485
\bibitem[Eggelton \& Kiseleva (1995)]{egg95}Eggelton P., Kiseleva L., 1995, \apj, 455, 640
\bibitem[Esposito et al.(2011)]{esp11}Esposito, S. et al. proc. 2011, SPIE 8149, 814902-10
\bibitem[Esposito et al.(2010a)]{esp10}Esposito, S. et al. proc. 2010a SPIE 7736, 773609-12
\bibitem[Esposito et al.(2010b)]{esp10a}Esposito, S. et al. proc. 2010b Applied Optics, 49, issue 31, p. G174
\bibitem[Esposito et al.(2012)]{esp12}Esposito, S. et al. 2012 proc. SPIE Volume 8447, article id. 84470U 
\bibitem[Fischer \& Marcy(1992)]{fis92}Fischer, D. A., Marcy, G. W. 1992, \apj, 396, 178
\bibitem[Freed, Close, \& Siegler(2003)]{fre02}Freed, M., Close, L.M., \& Siegler, N. 2003, \apj, 584, 453
\bibitem[Garrel, Guyon, \& Baudoz (2012)]{gar12}Garrel, Vincent; Guyon, Olivier; Baudoz, Pierre, PASP, 124, 861. 
\bibitem[Genzel \& Stutzki (1989)]{gen89}Genzel R., Stutzki J., 1989, ARA\&A 27, 41
\bibitem[Graves et al.(1998)]{gra98}Graves, J.E., Northcott, M.J., Roddier, F.J., Roddier, C.A., Close, L.M. 1988, Proc. SPIE Vol. 3353, p. 34-43. Adaptive Optical System Technologies, D. Bonaccini, R.K. Tyson, Eds.
\bibitem[Grellmann et al.(2013)]{gil13}Grellmann, R., Preibisch, T., Ratzka, T., Kraus, S. Helminiak, K.G., Zinnecker, H. 2013, A\&A 550, 82.
\bibitem[Gizis et al.(2003)]{giz03}Gizis, J.E. et al. 2003, \apj, 120, 1085
\bibitem[Hillenbrand \& Carpenter(2000)]{hil00}Hillenbrand L.A., \& Carpenter J. 2000 \apj, 540, 236
\bibitem[Hillenbrand \& Hartmann(1998)]{hil98}Hillenbrand L.A., \& Hartmann L.W. 1998 \apj, 492, 540
\bibitem[Hinz et al.(2002)]{hin02}Hinz J.L., McCarthy D.W., Simons, D.A., Henry T.J., Kirkpatrick J.D., McGuire P.C. 2002, \aj, 123, 2027
\bibitem[Hinz et al.(2010)]{hin10}Hinz P.M. et al. 2010, \apj 716 417
\bibitem[Hodapp et al.(1996)]{hod96}Hodapp, K.-W., Hora, J. L., Hall, D. N. B., Cowie, L. L., Metzger, M., Irwin, E., Vural, K., Kozlowski, L. J., Cabelli, S. A., Chen, C. Y., Cooper, D. E., Bostrup, G. L., Bailey, R. B., Kleinhans, W. E. 
1996, New Astronomy, 1, 177
\bibitem[Kiseleva \& Eggelton(1994)]{kis94} Kiseleva L.G., Eggelton P.P., Anosova J.P., 1994, MNRAS 267, 161
\bibitem[Kiseleva et al.(1994)]{kis94a} Kiseleva L.G., Eggelton P.P., Orlov V.V., 1994, MNRAS 270, 936
\bibitem[Kraus et al.(2007)]{kra07} Kraus, S. et al. 2007, A\&A 466, 649
\bibitem[Kraus et al.(2009)]{kra09} Kraus, S. et al. 2009, A\&A 497, 195
\bibitem[Kohler et al.(2006)]{koh06} Kohler, R. et al. 2006, A\&A 458, 461
\bibitem[Kopon et al.(2012)]{kop12} Kopon, D. et al. 2012, SPIE, Volume 8447, id. 84473D-84473D-15.  
\bibitem[Law et al.(2009)]{law09}Law, N. M.; Mackay, C. D.; Dekany, R. G.; Ireland, M.; Lloyd, J. P.; Moore, A. M.; Robertson, J. G.; Tuthill, P.; Woodruff, H. C. 2009, ApJ 692, 924.
\bibitem[Lloyd-Hart(2000)]{llo00}Lloyd-Hart M. 2000, PASP 112, 264
\bibitem[Males et al.(2012)]{mal12}Males, J.R. et al. 2012, SPIE, Volume 8447, id. 844742-844742-12.
\bibitem[Males et al.(2013)]{mal13}Males, J.R. et al. 2013 in prep.
\bibitem[McCaughrean \& Stauffer(1994)]{mcc94}McCaughrean M.J, \& Stauffer J.R., 1994, \aj, 108, 1382
\bibitem[McCaughrean (2000)]{mcc11}McCaughrean M.J 2000, The Formation of Binary Stars, Proceedings of IAU Symp. 200, held 10-15 April 2000, in Potsdam, Germany, Edited by Hans Zinnecker and Robert D. Mathieu, 2001, p. 169.
\bibitem[McCarthy \& Zuckerman(2004)]{mcc03}McCarthy C., Zuckerman, B. (2004), AJ 127, 2871 
\bibitem[McCarthy et al.(1998)]{mcc98}McCarthy D.W. et al. 1998, Proc. SPIE 3354 750
\bibitem[McDonald \& Clarke(1995)]{mcd95}McDonald, J. M., \& Clarke, C. J. 1993, \mnras, 262, 800
\bibitem[Menten et al.(2007)]{men07}Menten K. M.; Reid, M. J.; Forbrich, J.; Brunthaler, A. 2007, A\&A 474, 515. 
\bibitem[Morzinski et al.(2010)]{mor10}Morzinski, Katie; Johnson, Luke C.; Gavel, Donald T.; Grigsby, Bryant; Dillon, Daren; Reinig, Marc; Macintosh, Bruce A. 2010 Proceedings of the SPIE, Volume 7736, pp. 77361O-77361O-16
\bibitem[Morzinski et al.(2013)]{mor13}Morzinski, Katie; et al. 2013 in prep. 
\bibitem[Parker et al.(2011)]{par11}Parker, R.J., Goodwin, S., Allison, R.J. 2011, \mnras 418, 2565
\bibitem[Petr et al.(1998)]{pet98}Petr M.G., Du Foresto V., Beckwith S.V.W., Richichi A., McCaughrean M.J. 1998, \apj, 500, 825
\bibitem[Potter et al.(2002a)]{pot02a}Potter, D. et al. 2002a \apj, 567, 113
\bibitem[Reid et al.(2001a)]{rei01a}Reid, I. N., Gizis, J.E., Kirkpatrick, J.D., Koerner, D. W. 2001a, \aj, 121, 489
\bibitem[Reid et al.(2001b)]{rei01b}Reid, I. N., Burgasser, A. J., Cruz, K. L., Kirkpatrick, J. D., Gizis, J. E. 2001b, \aj, 121, 1710
\bibitem[Reipurth \& Clarke(2001)]{rep02} Reipurth, B. \& Clarke, C. 2001, \aj, 122, 432
\bibitem[Ricci et al.(2007)]{ric07}Ricci, Luca; Robberto, M.; Soderblom, D. R.; Kozhurina-Platais, V. 2007, BAAS 211, 8923
\bibitem[Schertl et al.(2003)]{sch03}Schertl, D., Balega, Y.Y., Preibisch, Th., \& Weigelt, G. 2003, A\&A, 402, 267
\bibitem[Siegler et al.(2003)]{sie02}Siegler, N., Close, L.M., Mamajek, E., Freed, M. 2003, \apj, 598, 1265
\bibitem[Siess Forestini \& Dougados(1997)]{sie97}Siess L., Forestini M., Dougados C., 1997, A\&A 324, 556
\bibitem[Simon, Close, \& Beck(1999)]{sim99} Simon, M., Close, L.M., \& Beck, T. 1999, \aj, 117, 1375
\bibitem[Sterzik \& Durisen(1998)]{ste98}Sterzik, M. F., \& Durisen, R. H. 1998 A\&A, 339, 95  
\bibitem[Stetson(1987)]{ste87}Stetson, P. B. 1987, \pasp, 99, 191
\bibitem[Thomas-Osip et al.(2010)]{tho10} Thomas-Osip Joanna E.; McCarthy, Patrick; Prieto, Gabriel; Phillips, Mark M.; Johns, Matt 2010,Proceedings of the SPIE, Volume 7733, pp. 77331L-77331L-10. 
\bibitem[Wainscoat \& Cowie(1992)]{wai92} Wainscoat R. J., \& Cowie, L.L. 1992, \aj, 103, 332.
\bibitem[Weigelt et al.(1999)]{wei99}Weigelt G., Balega, Y., Preibisch T., Schertl D., Scholler M., Zinnecker H. 1999, A\&A, 347, L15
\bibitem[Wildi et al.(2003a)]{wil03}Wildi F., Brusa G., Riccardi A., Lloyd-Hart M., Martin H.M., L.M. Close 2003, proc. SPIE 4839, 155
\bibitem[Wildi et al.(2003b)]{wil03b} Wildi F. Brusa G., Lloyd-Hart M., Martin H.M., L.M. Close, Riccardi A. 2003b, proc SPIE 5169, 17.



\end{thebibliography}
\end{document}